\documentclass{IEEEtran4PSCC}
\ifCLASSINFOpdf
   \usepackage[pdftex]{graphicx}
\else
   \usepackage[dvips]{graphicx}
\fi

\usepackage[cmex10]{amsmath}
\usepackage{amsfonts}
\usepackage{algorithm}
\usepackage{algpseudocode}

\usepackage{ntheorem}
\theoremseparator{:}
\newtheorem{hyp}{Hypothesis}

\usepackage{fnpct}
\ifCLASSOPTIONcompsoc \usepackage[caption=false,font=normalsize,labelfont=sf,textfont=sf]{subfig}
\else
\usepackage[caption=false,font=footnotesize]{subfig}
\fi
\usepackage{tikz}
\makeatletter
\let\old@ps@headings\ps@headings
\let\old@ps@IEEEtitlepagestyle\ps@IEEEtitlepagestyle
\def\psccfooter#1{%
    \def\ps@headings{%
        \old@ps@headings%
        \def\@oddfoot{\strut\hfill#1\hfill\strut}%
        \def\@evenfoot{\strut\hfill#1\hfill\strut}%
    }%
    \def\ps@IEEEtitlepagestyle{%
        \old@ps@IEEEtitlepagestyle%
        \def\@oddfoot{\strut\hfill#1\hfill\strut}%
        \def\@evenfoot{\strut\hfill#1\hfill\strut}%
    }%
    \ps@headings%
}
\makeatother

\begin{document}

\title{Model-less Robust Voltage Control in Active Distribution Networks using Sensitivity Coefficients Estimated from Measurements}

\author{
    \IEEEauthorblockN{
    Rahul Gupta$^1$, Fabrizio Sossan$^2$,  Mario Paolone$^1$\\}
    \IEEEauthorblockA{$^{1}$Distributed Electrical Systems Laboratory, EPFL, Switzerland, $^{2}$PERSEE, Mines ParisTech, France\\
    rahul.gupta@epfl.ch, fabrizio.sossan@mines-paristech.fr, mario.paolone@epfl.ch
   }}
    
\maketitle
\begin{abstract}
Measurement-rich power distribution networks may enable distribution system operators (DSOs) to adopt model-less and measurement-based monitoring and control of distributed energy resources (DERs) for mitigating grid issues such as over/under voltages and lines congestions. However, measurement-based monitoring and control applications may lead to inaccurate control decisions due to measurement errors. In particular, estimation models relying on regression-based schemes result in significant errors in the estimates (e.g., nodal voltages) especially for measurement devices with high Instrument Transformer (IT) classes. The consequences are detrimental to control performance since this may lead to infeasible decisions. This work proposes a model-less robust voltage control accounting for the uncertainties of measurement-based estimated voltage sensitivity coefficients. The coefficients and their uncertainties are obtained using a recursive least squares (RLS)-based online estimation, updated whenever new measurements are available. This formulation is applied to control distributed controllable photovoltaic (PV) generation in a distribution network to restrict the voltage within prescribed limits. The proposed scheme is validated by simulating a CIGRE low-voltage system interfacing multiple controllable PV plants. 
\end{abstract}
\section{Introduction}
Distribution system operators (DSOs) are required to operate their networks ensuring the quality of supply (QoS) while respecting the network's physical limits \cite{guide2004voltage, CIGREREF, IEEE_practice}. However, the progressive installation of decentralized generation such as Photo-voltaic (PV) units in distribution networks 
is causing power quality issues such as voltage violations \textcolor{black}{as well as} congestions in \textcolor{black}{both} lines and transformers. 
Conventional methods tackle these problems by passive curtailments of loads/generations, generation tripping, shunt capacitor banks \textcolor{black}{and, eventually expensive network reinforcement}. Authors in \cite{hatziargyrioucigre} lay out the potential ways to manage the electricity supply in modern power systems with a high amount of renewable generations. \textcolor{black}{In order to limit expensive grid reinforcement programs,} DSOs \textcolor{black}{may} need to adopt intelligent control \textcolor{black}{schemes} of distributed energy resources (DERs) (e.g., \cite{cigre2011c6, pilo2012planning}) for the safe operation of \textcolor{black}{their} grids.

Voltage control is one of the widely \textcolor{black}{acknowledged} control schemes \textcolor{black}{to be adopted and improved} in power distribution networks. Conventional voltage \textcolor{black}{controls are} based on volt-var schemes, where only the reactive power is controlled to regulate nodal voltages. However, as shown in \cite{christakou2015real}, \textcolor{black}{the sole} reactive power control might not be enough especially for \textcolor{black}{grids with} high R/X ratio \textcolor{black}{of branches longitudinal impedances,} 
\textcolor{black}{the control of} both active and reactive powers \textcolor{black}{may be needed.} 
In the literature, this type of control can be broadly categorized into two kinds. The first relies on the network model (network topology, branch, and shunt parameters). These methods are also referred to as model-based methods. For example, in \cite{agalgaonkar2013distribution, gupta2020grid}, \textcolor{black}{it is proposed a} distributed control of PV inverters for regulating the nodal voltage magnitudes in a distribution grid where the grid constraints are modeled using the admittance matrix of the network. However, in many cases, the network parameters are either unavailable, partially missing, or outdated. Thanks to the increasing adoption of monitoring systems such as smart meters in present distribution networks, measurement-based/data-driven/model-less control schemes can be an alternative. \textcolor{black}{This} leads to the second kind of voltage control scheme often referred to as measurement-based \textcolor{black}{schemes} \cite{christakou2015real, su2019augmented, carpita2019low, da2019data, nowak2020measurement}. These \textcolor{black}{schemes} are used for real-time voltage control where the network model is inferred from the measurements. 
However, in all the reported model-less and measurement-based methods, the control or the estimation problem \textcolor{black}{does} not consider uncertainty on the estimated grid models (e.g., estimated sensitivity coefficients) and may result in wrong \textcolor{black}{control} decisions. The uncertainty on the measurement-based estimated model comes from the measurement noise of the instrument transformers (ITs). 
As reported in \cite{zhang2017locally, zhang2017noise}, \textcolor{black}{the} estimated sensitivity coefficients suffer high biases due to measurement noise and \textcolor{black}{fluctuating} values due to collinearity in the measured data set. 

In this work, we propose a model-less robust voltage control that accounts for the uncertainty on the measurement-based estimated sensitivity coefficients ensuring safe and reliable operation of the distribution grid.
The work comprises \textcolor{black}{the estimation of} the sensitivity coefficients and their uncertainties and use them to provide robustness against the inaccuracies of measurement-based estimated grid models \cite{bertsimas2011theory}.
The \textcolor{black}{proposed} voltage control problem consists of two stages: in the first stage, an estimation problem is solved to estimate the voltage sensitivity coefficients and their uncertainties. In the second stage, we solve a robust voltage control problem accounting for the uncertainties on the estimated coefficients. 
Overall, the contributions of this \textcolor{black}{paper} are as follows:
\begin{itemize}
    \item  we investigate the effect of uncertainties of the \textcolor{black}{voltage sensitivity coefficients estimates adopted in} model-less voltage control of distributed PV generations \textcolor{black}{in a power distribution systems}. We show how this may result in voltage violations;
    \item \textcolor{black}{we formulate a robust voltage control problem using the measurement-based estimated sensitivity coefficients and their uncertainties;}
    \item we present a performance comparison of different estimation techniques \textcolor{black}{of} measurement-based estimations of sensitivity coefficients for the proposed robust control.
\end{itemize}

The \textcolor{black}{performance assessment is carried out on} the CIGRE LV \cite{CIGREREF} network interfacing multiple controllable PV units. First, we evaluate different techniques for estimating the uncertainties and the sensitivity coefficients, then the dominant method is chosen \textcolor{black}{to be coupled with} a robust control scheme. To show the effectiveness of the \textcolor{black}{proposed} robust formulation, we compare \textcolor{black}{it with non-robust} voltage control case when \textcolor{black}{uncertainties are} not considered. The performance is also benchmarked against model-based control.

The paper is organised as follows. Sec.~\ref{sec:prob_stat} presents the model-less robust control framework, Sec.~\ref{sec:estimation} presents different estimation methods for the estimation of the sensitivity coefficients, Sec.~\ref{sec:v_control} presents the voltage control problem and its robust reformulation. Sec.~\ref{sec:sim_results} describes the considered test-case and respective estimation and control results, and finally Sec.~\ref{sec:conclusion} concludes the work.

\section{Proposed Model-less Robust Voltage Control Framework}
\label{sec:prob_stat}
Let us consider a power distribution network equipped with measurement devices capable of providing \textcolor{black}{high throughout measurements} on nodal voltage magnitudes and active/reactive powers. Let $N_b$ be the number of non-slack buses and the set $\mathcal{N}^b = \{1,\dots, N_b\}$ \textcolor{black}{defining the} bus indices. 
The distribution network hosts multiple DERs (for example, PV generation units) that can be controlled \textcolor{black}{to provide} active and reactive power support to the grid.
The objective is to control DERs in real-time or quasi-real-time such that grid constraints are always respected. The parameters and topology of the network are not known, so model-based \textcolor{black}{controls could not be implemented}. 
The control scheme \textcolor{black}{solely} relies on a model-less scheme, where the grid constraints (such as nodal voltages, lines, and transformer power flows) are accounted by models estimated from measurements. Although the model-less framework is generic and can be applied for various control schemes, this work focuses on the voltage control problem where the DERs are controlled in real-time to \textcolor{black}{avoid or mitigate} voltage problems.

The model-less control framework consists of two \textcolor{black}{stages}: in the first \textcolor{black}{one}, measurements on voltages and active/reactive power magnitudes are used to estimate the voltage sensitivity coefficients; these are then used \textcolor{black}{by} the voltage control \textcolor{black}{stage}. We \textcolor{black}{compute the uncertainties of} the estimated sensitivity coefficients \textcolor{black}{in order to formulate a} robust voltage control problem. The uncertainties of the estimates are inferred using the inverse of the Fischer information matrix \cite{soderstrom1989system}.
\begin{figure}[!h]
    \centering
    \includegraphics[width = 2.8in]{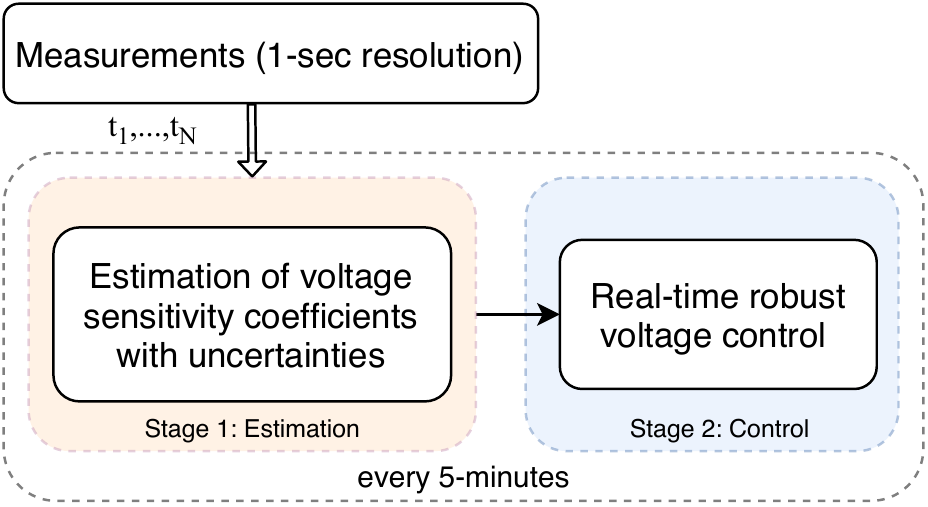}
    \caption{\textcolor{black}{Model-less/measurement-based robust voltage control framework}}
    \label{fig:flow}
\end{figure}
Figure~\ref{fig:flow} shows the flow diagram illustrating the proposed two-stage scheme for the robust model-less control framework. The first stage (on the left) is composed of a measurement-based estimation loop that stores nodal voltage magnitudes, active and reactive power measurements and estimates the voltage sensitivity coefficients and their uncertainties. Then, the block on the right solves the constrained optimization problem for controlling DERs to mitigate the voltage problems in the network. They are described as follows.
\section{Measurement-based Estimation of Voltage Sensitivity Coefficients and their Uncertainties}
\label{sec:estimation}
The voltage sensitivity coefficient of the $i-$th node with respect to absorbed/injected power at node $j$ is defined as 
\begin{align}
    K^P_{ij} = \frac{\partial V_i}{\partial P_j};~ K^Q_{ij} = \frac{\partial V_i}{\partial Q_j} \label{eq:Coeff_define}
\end{align}
where, $K^P_{ij}$ $K^Q_{ij}$ are the sensitivity coefficients of the nodal voltage magnitudes $V_i$ of node $i$ with respect to the active and reactive power injections $P_j, Q_j$ of node $j$.
Using the measurements on voltage magnitudes and active/reactive power injections, these sensitivity coefficients are estimated.
We assume following hypothesis \textcolor{black}{to hold true.} 
\begin{hyp}
    The DSO does not know the network parameters, and the topology, and the system state.
\end{hyp}
\begin{hyp}
    The distribution network is \textcolor{black}{equipped} with measurements units \textcolor{black}{providing} the operator with the measurements of voltage magnitudes, and active and reactive power injections at regular time intervals. The metering devices are aligned with a network time protocol (e.g. NTP \cite{mills1991internet}).
\end{hyp}
\begin{hyp}
    The sensitivity coefficients remain unchanged over a time window (5~minutes in this case) which is used to collect adequate number of measurements in the estimation problem.
\end{hyp}
\begin{hyp}
    The magnitude error from the ITs and voltmeter measurements is Gaussian. It behaves according to the standards and they do not have a bias. 
\end{hyp}

The objective is to estimate the voltage sensitivity coefficients and their uncertainties by using the measurements of nodal voltage magnitudes, active and reactive powers. The method is described as follows.
\subsection{Estimation model}
Using the coefficient definition in \eqref{eq:Coeff_define} and Taylor's first-order approximation, the magnitude deviation of the nodal voltages at time $t_k$ \textcolor{black}{for node $i$} can be written as
{\color{black}
\begin{align}
    & \underbrace{\Delta{V}_{i,t_k}}_{\gamma_{t_k}} \approx  \underbrace{[\Delta\mathbf{P}_{t_k} ~ \Delta\mathbf{Q}_{t_k}]}_{h_{t_k}}
    \underbrace{\begin{bmatrix}
        \mathbf{K}^P_{i,t_k}\\
        \mathbf{K}^Q_{i,t_k}
    \end{bmatrix}}_{\mathbf{X}}\label{eq:V_linear}
\end{align}
where 
${V}_{i,t_k} - {V}_{i,t_{k-1}} = \Delta{V}_{i,t_k} \in \mathbb{R}$ is the deviation of nodal voltage magnitude of $i-$th node, vectors $\mathbf{P}_{t_k} - \mathbf{P}_{t_{k-1}} = \Delta\mathbf{P}_{t_k}, \mathbf{Q}_{t_k} - \mathbf{Q}_{t_{k-1}} = \Delta\mathbf{Q}_{t_k} \in \mathbb{R}^{N_b}$ include deviations of active and reactive powers of all the nodes from timestep $t_{k-1}$ to $t_k$. 
The vectors $\mathbf{K}^P_{i,t_k}, \mathbf{K}^Q_{i,t_k} \in \mathbb{R}^{N_b}$ include voltage sensitivity coefficients of $i-$th node with respect injections of nodes $j\in\mathcal{N}^b$.
It should be noted that the approximation in \eqref{eq:V_linear} of the power-flow equations involves two errors: (i) the linearization and (ii) the measurement noise. In this work, we assume that the linearization error is negligible compared to the one due to the measurement noise. This assumption is reasonable if the state of the system is slow varying and the control is acting in quasi real-time. 
Assuming that we have measurements for time $t = t_1 \dots, t_N$ and coefficients do not change for $N$ time-steps (Hypothesis 3), Eq. \eqref{eq:V_linear} can be written as 
\begin{align}
    \mathbf{\Gamma} \approx \mathbf{H}\mathbf{X} \label{eq:Lmodel}
\end{align}
where, $\mathbf{{\Gamma}} \in \mathbb{R}^N = [\gamma_{t_1} \gamma_{t_2} \dots \gamma_{t_N}]^T$, $ \mathbf{H} \in \mathbb{R}^{N \times 2N_b} = [h_{t_1} h_{t_2} \dots h_{t_N}]^T$ and $\mathbf{X} \in R^{2N_b}$ includes $\mathbf{K}^P_{i,t_k}$ and $\mathbf{K}^Q_{i,t_k}$.
Eq. \eqref{eq:Lmodel} can be re-written assuming noise model to be white Gaussian (Hypothesis 4).}
\begin{align}
    \mathbf{\Gamma} = \mathbf{H}\mathbf{X} + \boldsymbol{\mathcal{W}} && \boldsymbol{\mathcal{W}} \in \mathcal{N}(\mathbf{0}, \Sigma)\label{eq:LmodelwNoise},
\end{align}
$\Sigma$ refers to the noise covariance matrix.
\subsection{Estimation technique}\label{sec:estimation_tech}
\begin{figure}[h]
    \centering
    \includegraphics[width=2.7in]{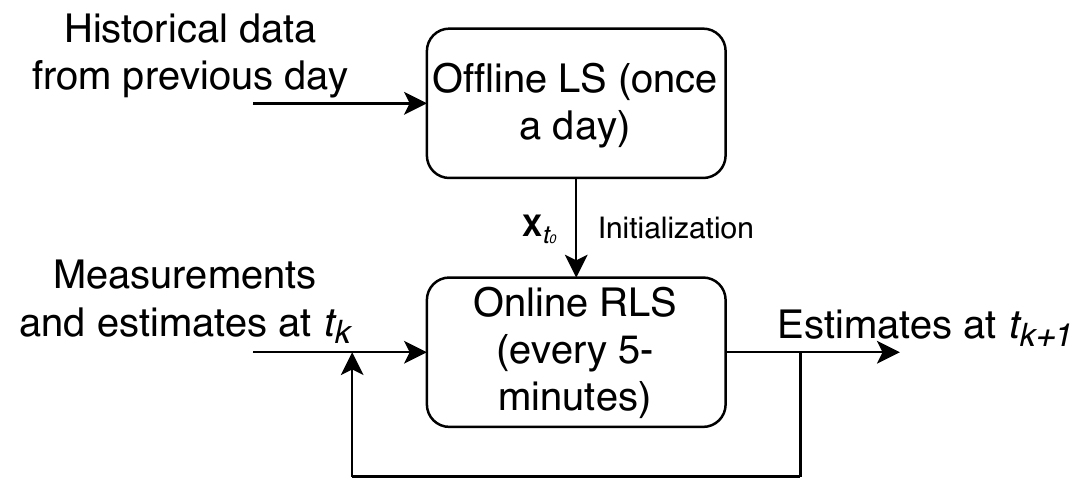}
    \caption{Flow diagram for two-stage estimation of sensitivity coefficients.}
    \label{fig:est_flow}
\end{figure}

The linear model in \eqref{eq:LmodelwNoise} is typically solved for $\mathbf{X}$ by minimizing the norm-2 difference of the residual, known as Least-Squares (LS).
However, the LS method does not perform well in case of \textcolor{black}{low excitation (nodal power injections are low) and suffers from the problem of multicollinearity (power injections at different nodes are very similar)} \cite{da2019data, zhang2017noise}.
Also, the sensitivity coefficients vary as a function of network's states \textcolor{black}{so,} it is necessary to use the most recent estimates during a real-time control. Thus, an online estimation scheme was used in \cite{da2019data, nowak2020measurement} that used recursive least square (RLS)-based estimation \textcolor{black}{coupled with an offline} LS. In this work, we use this scheme to estimate the sensitivity coefficients. Figure~\ref{fig:est_flow} shows the \textcolor{black}{dataflow of the} estimation \textcolor{black}{process}.
First, the LS is used to get a rough estimates of the coefficients. \textcolor{black}{Then, the} RLS is used to refine the LS estimates by using the latest information on the voltage and power measurements. The LS is solved off-line using a large number of historical measurements. The RLS problem is solved at each time step using recent measurements \textcolor{black}{where the LS estimation is used to initialize the RLS. Both the processes are described next.}

\subsubsection{Offline LS}
Offline LS problem is formulated as
\begin{align}\label{eq:RegularisedLS}
    \widehat{\mathbf{X}} = \underset{\mathbf{X}}{\text{min}}||\mathbf{\Gamma} - \mathbf{H}\mathbf{X}||_2 + \lambda\mathbf{X}^T\mathbf{X}
\end{align}
where $\lambda$ is a positive number that serves as \textcolor{black}{a} regularization parameter and is used to avoid ill-conditioned information matrix (i.e., in case of multi-collinearity nodal injections). 
 \textcolor{black}{The} closed-form solution of \eqref{eq:RegularisedLS} is obtained in view of its quadratic and unconstrained nature as,
\begin{align}
    & \widehat{\mathbf{X}}_{t_0} = (\mathbf{H}^T\mathbf{H} + \lambda \mathbf{I})^{-1}\mathbf{H}^T\mathbf{\Gamma} = (\mathbf{R}_{t_0}+ \lambda \mathbf{I})\mathbf{H}^T\mathbf{\Gamma} 
    \label{eq:LSsol}
\end{align}
where $\mathbf{I}$ is the identity matrix. The covariance matrix is defined as inverse of the information matrix, i.e. $\mathbf{P}^\text{cov}_{t_0} = \mathbf{R}_{t_0}^{-1}$.
\subsubsection{Online RLS}
In this scheme, an online recursive estimation is performed using the most recent measurements. It utilizes the estimates from the previous time step and measurements at the current time step. 
RLS updates the estimates whenever the new data is available. LS solution in \eqref{eq:LSsol} is used \textcolor{black}{to initialize the RLS stage}. The use of exponential forgetting \textcolor{black}{factor applied to the observations} is advised to give less importance to previous measurements \cite{parkum1992recursive}. The forgetting factor $0 < \mu \leq 1$ is reflected in the covariance matrix update.
\begin{align}
     & \mathbf{R}_{t_k} = \mu\mathbf{R}_{t_{k-1}} + h_{t_k}^Th_{t_k}
\end{align}
This results in  \textcolor{black}{the} following iterative updates. \label{eq:RLS-F}
\begin{subequations}
\begin{align}
        & e_{t_k} = \gamma_{t_k} - h_{t_k}\widehat{\mathbf{X}}_{t_{k-1}}\\
        & \widehat{\mathbf{X}}_{t_k} = \widehat{\mathbf{X}}_{t_{k-1}} + \mathcal{L}_{t_k}e_{t_k}\\
        & \mathcal{L}_{t_k} = \frac{\mathbf{P}^\text{cov}_{t_{k-1}}h^T_{t_k}}{\mu + h_{t_k}\mathbf{P}^\text{cov}_{t_{k-1}}h^T_{t_k}}\\
        & \mathbf{P}^\text{cov}_{t_{k}} = (\mathbf{I} -  \mathcal{L}_{t_k}h_{t_k})\mathbf{P}^\text{cov}_{t_{k-1}}/\mu \label{eq:RLSFd}
\end{align}
\end{subequations}
where, $ \mathcal{L}$  \textcolor{black}{is} the estimated gain and $e$ the residual. \textcolor{black}{In the following,} this scheme is referred  \textcolor{black}{to as} RLS-F.

\textcolor{black}{As reported in \cite{parkum1992recursive, vahidi2005recursive},} the RLS-F scheme suffers from the windup problem of the covariance matrix . It may lead to very large covariances resulting in \textcolor{black}{large} estimates variances. Multiple schemes are proposed in the literature to solve this problem. They are briefly described \textcolor{black}{next}.
\paragraph{Constant-Trace scheme (RLS-CT)} in \cite{fortescue1981implementation}, \textcolor{black}{it is discussed how to limit} the windup problem of \textcolor{black}{the} co-variance matrix by setting an upper bound on the trace sum of the covariance matrix and adding an identity matrix $\mathbf{I}$. The scheme uses two different factors $c_1$ and $c_2$ such that $c_1/c_2 = 10e3; ~ h_{t_k}h_{t_k}^T c_1 >> 1 $.
The covariance matrix is modified as:
\begin{align}
        & \mathbf{P}^\text{cov}_{t_{k}} = c_1\mathbf{P}^\text{cov}_{t_{k}}/{\text{trace}}(\mathbf{P}^\text{cov}_{t_{k}}) + c_2\mathbf{I}
\end{align}
\paragraph{Selective forgetting (RLS-SF)}
in \cite{parkum1992recursive} \textcolor{black}{it is proposed} to use selective forgetting factor, i.e., to use different forgetting factors for different eigenvalues of the covariance matrix. These forgetting factors are computed and updated iteratively to limit the windup problem of the covariance matrix. The gain and covariance matrix \textcolor{black}{are} updated as follows. 
\begin{subequations}
\begin{align}
       & \mathcal{L}_{t_k} = \frac{\mathbf{P}^\text{cov}_{t_{k-1}}h^T_{t_k}}{1 + h_{t_k}\mathbf{P}^\text{cov}_{t_{k-1}}h^T_{t_k}}\\
        & \mathbf{P}^\text{cov}_{t_{k}} = \sum_{i=1}^{2N_b}\frac{\tau_{i,t_k}}{\mu_i}u_{i,t_k}^Tu_{i,t_k} \label{eq:SFe}.
\end{align}
Here, $u_{i,t_k}$ denotes the eigenvectors of $\mathbf{P}^\text{cov}_{t_k}$ in Eq. \eqref{eq:RLSFd} and $ \tau_{i,t_{k}}$ \textcolor{black}{the corresponding} eigenvalues. \textcolor{black}{It proposed to limit $\tau_{i,t_{k}}$ by} a function $f$ that keeps it within bounds $[\tau_{\text{min}}~ \tau_{\text{max}}]$:
\begin{align}
        & \tau_{i,t_{k}} = f( \tau_{i,t_{k-1}}) \label{eq:SFf}\\
        & f(x) = 
        \begin{cases}
                x, & x>\tau_{\text{max}}\\
                \tau_{\text{min}} + (1 - \tau_{\text{min}}/{\tau_{\text{max}}})x & x \leq \tau_{\text{min}}
        \end{cases} \label{eq:SFg}
\end{align}
\end{subequations}
More information on the tuning of RLS-SF is in \cite{parkum1992recursive} and \cite{fortescue1981implementation}.

\paragraph{Directional forgetting (RLS-DF)} in \cite{cao1999novel,bittanti1990exponential} proposed directional forgetting algorithm where the matrix $\mathbf{R}$ is decomposed into two parts: the first part is fully propagated to the next time step, whereas the second part is propagated with a forgetting factor, $\mu$. This method was first proposed in \cite{cao1999novel} and termed as ``directional forgetting" as the two parts of the gain matrix are orthogonal to each other. 
Theoretical development supporting this algorithm is in \cite{bittanti1990exponential}. The iterative updates of RLS-DF are
\begin{subequations}
\begin{align}
        & \mathcal{L}_{t_k} = {\mathbf{P}}^{\text{cov}}_{t_k}h^T_{t_k}\\
        & \bar{\mathbf{P}}^{\text{cov}}_{t_{k-1}} = {\mathbf{P}}^{\text{cov}}_{t_{k-1}} + \frac{1-\mu}{\mu}\frac{h_{t_{k}}^T h_{t_k}}{h_{t_k}\mathbf{R}_{t_{k}}h^T_{t_k}}\\
        & {\mathbf{P}}^{\text{cov}}_{t_{k}} =  \bar{\mathbf{P}}^{\text{cov}}_{t_{k-1}} - \frac{\bar{\mathbf{P}}^{\text{cov}}_{t_{k-1}} h_{t_k}^T h_{t_k}\bar{\mathbf{P}}^{\text{cov}}_{t_{k-1}}}{1+ h_{t_k}\bar{\mathbf{P}}^{\text{cov}}_{t_{k-1}}\ h^T_{t_k}}\\
        & \mathbf{R}_{t_k} = [\mathbf{I} - \mathbf{M}_{t_k}]\mathbf{R}_{t_{k-1}} + h_{t_k}^T h_{t_k}\\
        & \mathbf{M}_{t_k} = (1-\mu) \frac{\mathbf{R}_{t_{k-1}} h_{t_k}^T h_{t_k}}{h_{t_k}\mathbf{R}_{t_{k-1}}h^T_{t_k}}
\end{align}
\end{subequations}

The updates strategies for the covariance matrix directly affects the estimates and \textcolor{black}{their} uncertainties. \textcolor{black}{The numerical performance of these schemes (i.e. RLS-F, RLS-CT, RLS-SF and RLS-DF) are assessed in the} results section.
\subsubsection{Estimation of uncertainties on sensitivity coefficients}
In this work, we propose \textcolor{black}{to account for the uncertainties of} the estimated sensitivity coefficients to robustify the voltage control. The \textcolor{black}{uncertainties} are computed using the co-variance matrix \textcolor{black}{given by}
\begin{align}
    \sigma_{\mathbf{X}} = \sigma_r\sqrt{\text{diag}(\mathbf{P}^\text{cov})}
\end{align}
where $\sigma_r$ is the estimated standard deviation of residuals inferred post-estimation. They are continuously updated during the RLS estimation stage.
The uncertainty on the estimated coefficients are estimated to be $\pm 3 \sigma_{\mathbf{X}}$ corresponding to the 99~\% confidence interval.
\section{Model-less Robust Voltage Control Problem}\label{sec:v_control}
\textcolor{black}{As previously mentioned, the  proposed control} uses the estimated coefficients and their uncertainties to formulate a robust voltage control problem. The robustification uses the technique \textcolor{black}{from \cite{christakou2017voltage} where the robustness of a model-based voltage control was formulated against uncertainty in the resistances of the grid's branch impedances}. In contrast, in this work we propose a measurement-based and mode-less way to define a robust voltage control.
\subsection{Voltage control problem without considering uncertainty on the estimates (Non-robust)}
\textcolor{black}{Let us} consider a distribution network connected with controllable PV generation units such that \textcolor{black}{their} active and reactive power injections can be controlled. Let the set $\mathcal{N}^{\text{pv}}$ includes indices of the PV units. The objective is to control active/reactive power injections ($P_{j,t_k}, Q_{j,t_k}, j\in\mathcal{N}^{\text{pv}}$) such that the nodal voltages are within the statutory bounds.
Additionally, the local objective of the PV units is to minimize the curtailment of \textcolor{black}{their} active power generation and provide reactive power support. \textcolor{black}{ The problem we solve at time $t_k$ is to minimize curtailments of PV plants:}
\begin{subequations}\label{eq:non-Robust}
\begin{align}
    \underset{{P}_{j,t_k}, {Q}_{j,t_k}, \forall j\in\mathcal{N}^b} {\text{minimize}}~ \sum_{j\in\mathcal{N}^{\text{pv}}}\Big\{(P_{j,{t_k}}-\widehat{P}_{j,t_k})^2  + (Q_{j,t_k})^2 \Big\} 
\end{align} 
subject to
the constraint on the PV generation limited by short-term MPP forecast $\widehat{P}_{j,t_k}$, 
\begin{align} 
    & 0 \leq P_{j,t_k} \leq  \widehat{P}_{j,t_k} \label{eq:PVp_limit} && j \in \mathcal{N}^\text{pv}
\end{align}
the capability constraint of the converter rating ${S}_j^{\text{max}}$,
\begin{align}
    & 0 \leq (P_{j,t_k})^2 + (Q_{j,t_k})^2 \leq  ({S}_j^{\text{max}})^2 && j \in \mathcal{N}^\text{pv},  \label{eq:capability_PV}
\end{align}
\textcolor{black}{and} the minimum power factor constraint
\begin{align}
    &  Q_{j,t_k} \leq P_{j,t_k}\zeta && j \in \mathcal{N}^\text{pv}\label{eq:pf1}\\
    &  -Q_{j,t_k}  \leq P_{j,t_k}\zeta && j \in \mathcal{N}^\text{pv}\label{eq:pf2}.
\end{align}
Here, $\zeta = \sqrt{(1-\text{PF}^2_{\text{min}})/\text{PF}^2_{\text{min}}}$,  \textcolor{black}{being} $\text{PF}_{\text{min}}$ the minimum power-factor allowed for the PV operation  \textcolor{black}{of each PV plant.}
The final constraints are on the voltage \textcolor{black}{magnitudes,} which are bounded by [${V}^{\text{min}}, {V}^{\text{max}}$]. The voltage magnitudes are modeled by the estimated voltage sensitivity coefficients as
\textcolor{black}{
\begin{align}
\begin{aligned}
    {V}^{\text{min}} \leq {V}_{i,t_{k-1}} +  & \widehat{\mathbf{K}}^P_{i,t_{k-1}}\Delta{\mathbf{P}_{t_k}}+ \\ 
    & \widehat{\mathbf{K}}^Q_{i,t_{k-1}}\Delta{\mathbf{Q}_{t_k}} \leq {V}^{\text{max}} && \forall i \in \mathcal{N}^b \label{eq:volt_const} 
\end{aligned}
\end{align}}
The voltage sensitivity coefficients, $ \widehat{\mathbf{K}}^P_{i,t_{k-1}},  \widehat{\mathbf{K}}^Q_{i,t_{k-1}}$, are estimated online using one of the estimation scheme described in Sec.~\ref{sec:estimation} utilizing latest measurements on voltages and power magnitudes. 

As described earlier, the non-robust problem in \eqref{eq:non-Robust} does not account for the uncertainty on the estimates caused by measurement noise which might result in inaccurate control decisions leading to voltage violations.

\subsection{Robust voltage control problem}
We \textcolor{black}{here illustrate the} robust voltage control by accounting for the uncertainty on the measurement-based estimated voltage sensitivity coefficients. \textcolor{black}{The robust counterpart of \eqref{eq:non-Robust}} can be formulated by adding following constraints to \eqref{eq:non-Robust}
\begin{align}
    & \mathbf{K}^P_{i,{t_k}} \in [\widehat{\mathbf{K}}^P_{i,{t_k}}- \Delta \mathbf{K}^P_{i,{t_k}}, ~\widehat{\mathbf{K}}^P_{i,{t_k}}+ \Delta \mathbf{K}^P_{i,{t_k}}] \label{eq:v1} && \forall i \in \mathcal{N}^b\\
    & \mathbf{K}^Q_{i,{t_k}} \in [\widehat{\mathbf{K}}^Q_{i,{t_k}}- \Delta \mathbf{K}^Q_{i,{t_k}}, ~\widehat{\mathbf{K}}^Q_{i,{t_k}}+ \Delta \mathbf{K}^Q_{i,{t_k}}] && \forall i \in \mathcal{N}^b \label{eq:v2}.
\end{align}
\end{subequations}
Here, $\Delta \mathbf{K}^P_{i,t_k}, \Delta \mathbf{K}^Q_{i,t_k}$ be the estimated uncertainty on $\widehat{\mathbf{K}}^P_{i,t_k}, \widehat{\mathbf{K}}^Q_{i,{t_k}}$.
\textcolor{black}{As known,} accounting for the interval constraints makes \textcolor{black}{the problem} non-tractable in its original form. Thus, \textcolor{black}{it} is reformulated using the technique proposed in \cite{christakou2017voltage, bertsimas2004price} summarized hereafter.

We introduce auxiliary variables $z_i, g_{ij}, y^p_j, y^q_j, j\in \mathcal{N}^\text{pv}, i \in \mathcal{N}^b$. We also introduce a parameter $\Omega_i \in [0, |\mathcal{N}^\text{pv}|]$ which provides a trade-off between the robustness and conservativeness of the solution as described in \cite{christakou2017voltage}. 
Considering these auxiliary variables and following the robust \textcolor{black}{quadratic program  with linear constraints} in \cite{bertsimas2004price}, the robust counterpart of the problem can be formulated as
\begin{subequations}\label{eq:RobustVcontrol}
\begin{align}
    \underset{{P}_{j,t_k}, {Q}_{j,t_k}, \forall j\in\mathcal{N}^b} {\text{minimize}}~ \sum_{j\in\mathcal{N}^{\text{pv}}}\Big\{(P_{j,{t_k}}-\widehat{P}_{j,t_k})^2  + (Q_{j,t_k})^2 \Big\} 
\end{align} 
subject to:
\begin{align}
    \eqref{eq:PVp_limit}, \eqref{eq:capability_PV},
    \eqref{eq:pf1}, \eqref{eq:pf2}
\end{align}
\textcolor{black}{With the help of the auxiliary variables}, the constraints on the nodal voltages are reformulated as follows.
\begin{align}
& \begin{aligned}
    {V}_{i,t_{k-1}} + & \widehat{\mathbf{K}}^P_{i,t_{k-1}}\Delta{\mathbf{P}_{t_k}} + 
    \widehat{\mathbf{K}}^Q_{i,t_{k-1}}\Delta{\mathbf{Q}_{t_k}} + \\ 
    & z_i\Omega_i + \sum_{j\in \mathcal{N}^\text{pv}}g_{ij} \leq V^{\text{max}} ~~ \forall i \in \mathcal{N}^b\\
\end{aligned}\\
& \begin{aligned}
    {V}_{i,t_{k-1}} + &  \widehat{\mathbf{K}}^P_{i,t_{k-1}}\Delta{\mathbf{P}_{t_k}} + 
    \widehat{\mathbf{K}}^Q_{i,t_{k-1}}\Delta{\mathbf{Q}_{t_k}} - \\ 
    &  z_i\Omega_i - \sum_{j\in \mathcal{N}^\text{pv}}g_{ij} \geq V^{\text{min}} ~~ \forall i \in \mathcal{N}^b\\
\end{aligned}
\end{align}
\begin{align}
    & -{y}^p_j \leq \Delta{{P}_{j,t_k}} \leq {y}^p_j &&  \forall j \in \mathcal{N}^{\text{pv}}\\
    & -{y}^q_j \leq \Delta{{Q}_{j,t_k}} \leq {y}^q_j &&  \forall j \in \mathcal{N}^{\text{pv}}\\
    & z_i + g_{ij} \geq \Delta{K}_{ij, t_k}^P y^p_j && i \in \mathcal{N}^b, j\in \mathcal{N}^{\text{pv}}\\ 
    & z_i + g_{ij} \geq \Delta{K}_{ij, t_k}^Q y^p_j && i \in \mathcal{N}^b, j\in \mathcal{N}^{\text{pv}}\\ 
    & {y}^p_j, {y}^q_j, {z}_i, {g}_{ij} \geq 0  && i \in \mathcal{N}^b, j\in \mathcal{N}^{\text{pv}}.
\end{align}
\end{subequations}

The robust problem in \eqref{eq:RobustVcontrol} has a quadratic objective and linear constraints, hence it is convex and can be efficiently solved with any off-the-shelf solvers.


\section{Simulation and Results}\label{sec:sim_results}
\subsection{Test-case and input data}
\begin{figure}
    \centering
    \includegraphics[width=3.4in]{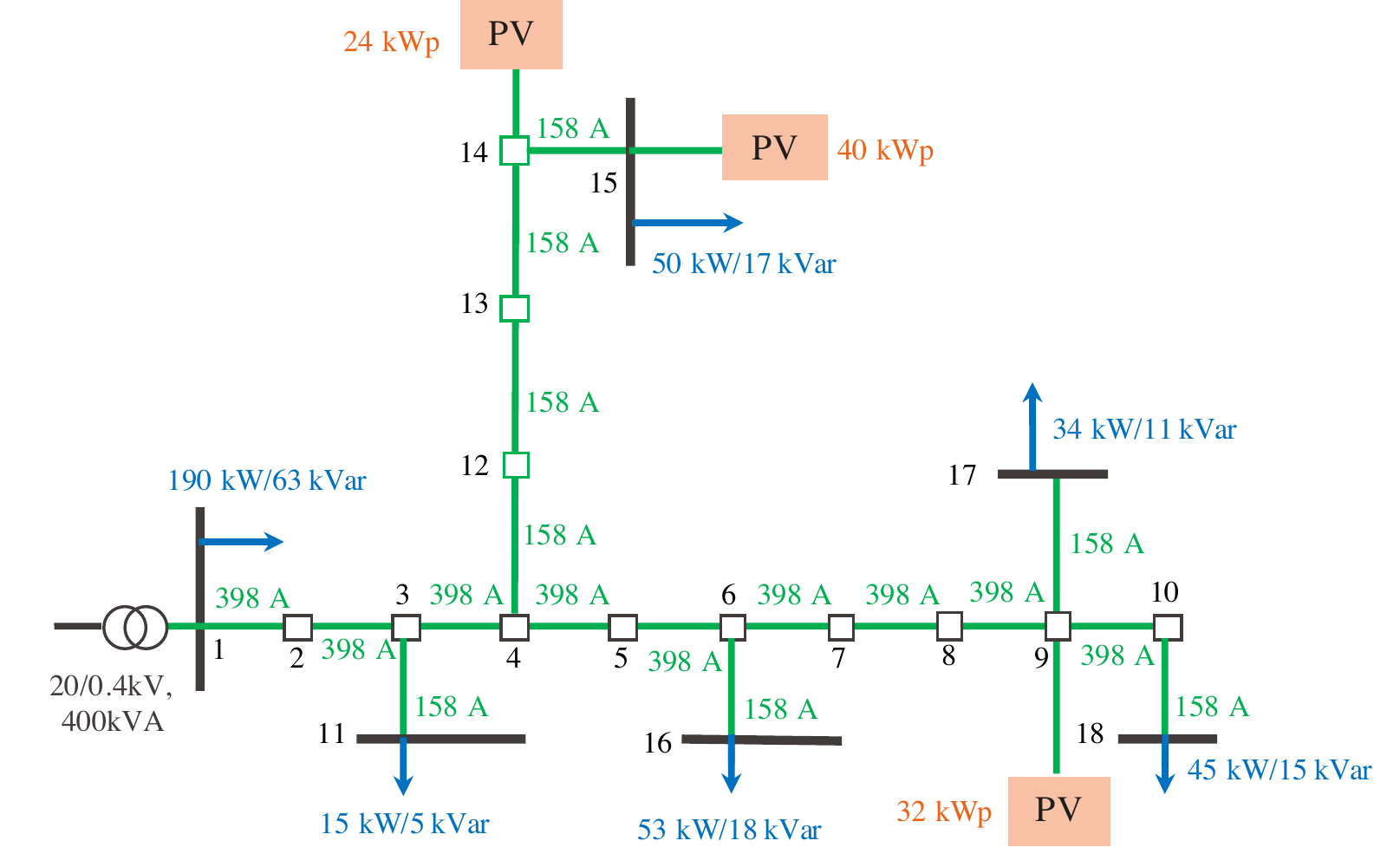}
    \caption{Topology of the CIGRE low-voltage system with distributed PV units.}
    \label{fig:network}
\end{figure}

For the validation of the measurement-based estimation and model-less control scheme and \textcolor{black}{the corresponding} performance evaluation, we consider a CIGRE benchmark low-voltage network \cite{CIGREREF}. The network is 0.4kV/400V 3-ph balanced system as shown in Fig.~\ref{fig:network}. 
\textcolor{black}{The nominal demands and the PV generation sites and sizes are also shown in the figure.}
In this case study, we assume reduced load condition such that the PV generation is causing over-voltages during the middle of the day.
\begin{figure}[h]
\centering
\subfloat[]{\includegraphics[width=0.95\columnwidth]{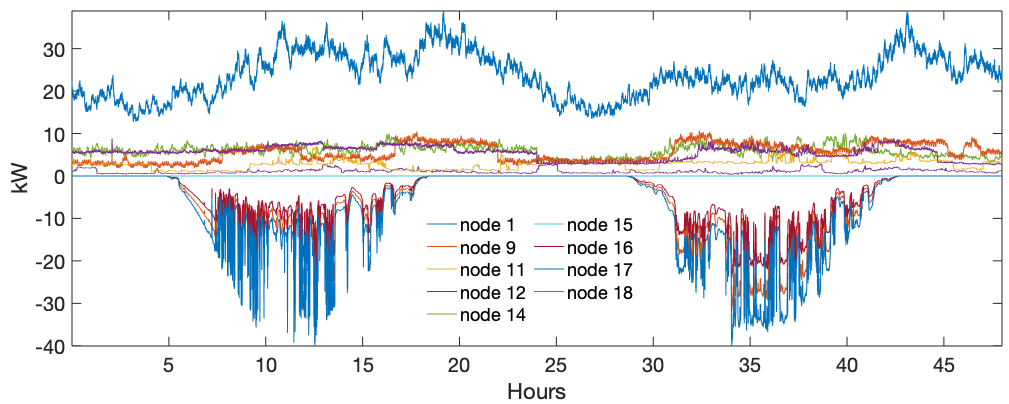}
\label{fig:Pplot}}
\hfil
\subfloat[]{\includegraphics[width=0.95\columnwidth]{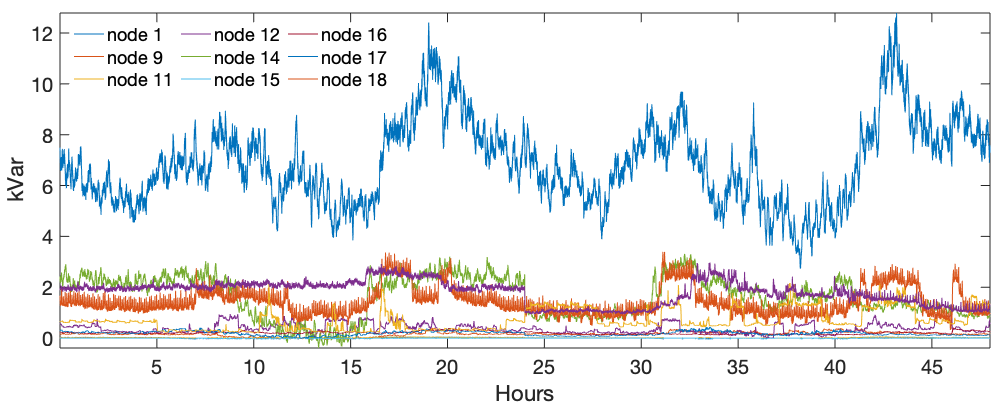}
\label{fig:Qplot}}
\caption{(a) Nodal active (in kW) and (b) reactive (in kVar) power injections for non-zero injection nodes.} \label{fig:PQplot}
\end{figure}
Figure~\ref{fig:PQplot}(a-b) shows the nodal active and reactive power injections. To obtain the ground-truth measurements of the voltage magnitudes and power injections, we carry simulated experiments performing load-flows by knowing the true admittance matrix of the grid. Then, the currents and voltages are corrupted with measurement noises characterized by the IT's specification described in \cite{IT_V, IT_C}. This process is described in \textbf{Algorithm 1}. The algorithm introduces noise \textcolor{black}{in polar coordinates (i.e.,} magnitudes and phase noise) on the voltage and currents, which is then used to compute the corrupted nodal active ($\mathbf{\tilde{P}}$) and reactive ($\mathbf{\tilde{Q}}$) power magnitudes. The specifications of the ITs are listed in Table~\ref{tab:instrument_class}.

\begin{table}[ht]
\caption{Errors specifications for different class of Instrument Transformers Defined by \cite{IT_V, IT_C}.}
\begin{center}
\begin{tabular}{ |c|c|c|c|c|} 
\hline
\textbf{IT class} & \multicolumn{2}{|c|}{\textbf{Voltage transformers}} & \multicolumn{2}{|c|}{\textbf{Current transformers}}  \\
\hline
\textbf{} & \textbf{mag. error} & \textbf{phase error} & \textbf{mag. error} & \textbf{phase error}  \\
 & (${\sigma^m}$) {[\%]} &  $(\sigma^p)$ {[rad.]} &  (${\sigma^m}$) {[\%]} &  $(\sigma^p)$ {[rad.]} \\
  \hline
         0.2 & 0.2 & 3e-3 & 0.2 & 3e-3 \\
         0.5 & 0.5 & 6e-3 & 0.5 & 9e-3\\
         1 & 1 & 12e-3 & 1 & 18e-3\\
   \hline
\end{tabular}\label{tab:instrument_class}
\end{center}
\end{table}

\begin{algorithm}
\caption{Raw-data generation}\label{alg:GenData}
\begin{algorithmic}[1]
\Require {Admittance: $\mathbf{Y}$, nodal power injections: $\mathbf{P}, \mathbf{Q}$}
\Procedure{GenData}{}
\For{$t_k=t_1:t_N$}
\State $[\mathbf{V}(t_k), \mathbf{I}(t_k)]$ = LoadFlow($\mathbf{P}(t_k), \mathbf{Q}(t_k)$, $\mathbf{Y}$)
\State $[\mathbf{\tilde{V}}(t_k), \mathbf{\tilde{I}}(t_k)]$ =
    \For{$\beta = [\mathbf{V}(t_k), \mathbf{I}(t_k)]$}
        \State $\delta^m = \mathcal{N}(0, {\sigma^m}|\beta|/3)$
        \State $|\beta| = |\beta| + \delta^m$
        \State $\delta^{p} = \mathcal{N}(0, {\sigma^p}/{3})$
        \State $\text{arg}(\beta) = \text{arg}(\beta) + \delta^{m}$
        \State $\beta = |\beta|\text{exp}(j~\text{arg}(\beta))$
    \EndFor
\State $\mathbf{\tilde{P}}(t_k)+ j\mathbf{\tilde{Q}}(t_k)$ = $\mathbf{\tilde{V}}(t_k)\mathbf{\tilde{I}}(t_k)^*$
\EndFor
\EndProcedure
\end{algorithmic}
\end{algorithm}

\subsection{\textcolor{black}{Performance} metrics}
\textcolor{black}{This section defines the metrics used in the performance assessment. The first metric is the classical}
root-mean-square-error (RMSE), defined as
\begin{align}
    \text{RMSE}(\mathbf{\mathbf{\widehat{X}}}) = \frac{||\mathbf{X}^{\text{true}} - \mathbf{\widehat{X}}||_2}{||\mathbf{X}^{\text{true}}||_2}.
\end{align}
where, $\mathbf{X}^{\text{true}}, \mathbf{\widehat{X}}$ are the true and estimated values of a generic quantity.

For the performance comparison on the estimation of the uncertainty intervals, we use metrics  \textcolor{black}{inspired by} \cite{khosravi2013prediction}:
the first is the prediction interval coverage probability (PICP) that counts the number of instances realization falling within the uncertainty bounds for a given confidence interval $\alpha$. It is
\begin{align}
    & \text{PICP} = \frac{1}{N}\sum_{t_k=t_1}^{t_N}b_{t_k}\\
    & b_{t_k} = \begin{cases}
            1 & \widehat{K}^P_{ij,t_k} - \Delta{K^P_{ij,t_k}}  \leq  \widehat{K}^P_{ij,t_k} \leq {K}^P_{ij,t_k} + \Delta{K^P_{ij,t_k}}  \\
            0 & \text{otherwise}.
    \end{cases}
    \end{align}
The second is \textcolor{black}{the} prediction interval normalized average width (PINAW):
\begin{align}
          \text{PINAW} = \frac{1}{N (K^P_{ij,\text{max}})}\sum_{t_k=t_1}^{t_N} (2\Delta{K^P_{ij,t_k}}).
\end{align}
\textcolor{black}{Being} $K^P_{ij,\text{max}}$ the maximum value of the coefficient in the series.
The final metric is \textcolor{black}{the} coverage width-based criterion (CWC), which quantifies the trade-off between high PICP and small PINAW.
\begin{align}
    & \text{CWC} = \text{PINAW}(1+\eta \text{(PICP)} e^{-(\nu(\text{PICP} - \alpha))}\\
    & \eta = \begin{cases}
            0, & \text{PICP} \leq \alpha \\
            1, & \text{otherwise}
    \end{cases}
\end{align}
The parameter $\nu$ can be set based on a tradeoff between the interval width penalization. We chose it to be $\nu = 50$. The \textcolor{black}{considered} confidence $\alpha$ is 99\%. 
\begin{figure*}[h]
\centering
\subfloat[$K^P_{15,15}$]{\includegraphics[width=0.66\columnwidth]{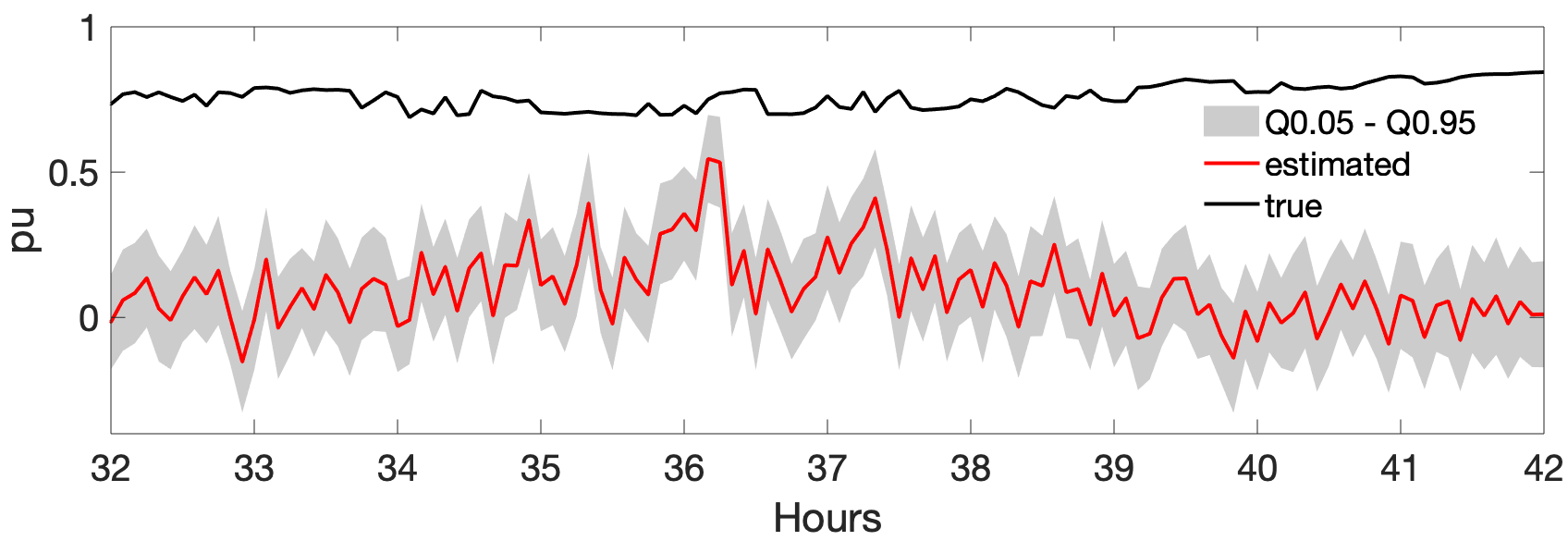}
\label{fig:LSa}}
\subfloat[$K^P_{14,8}$]{\includegraphics[width=0.66\columnwidth]{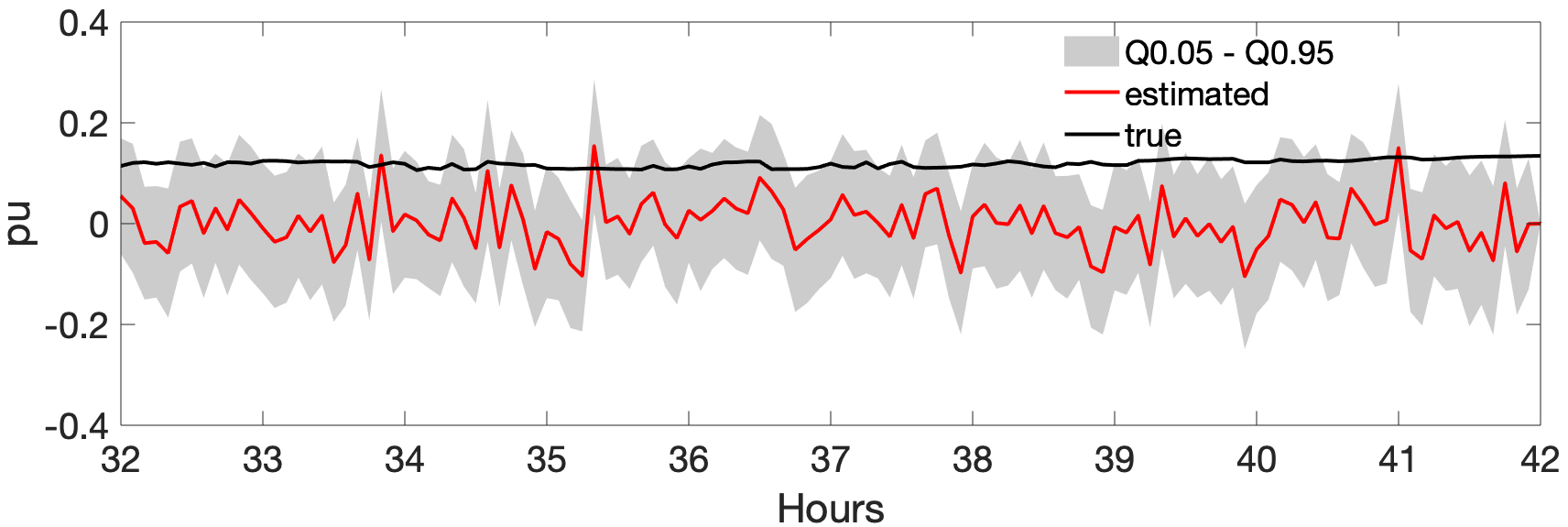}
\label{fig:LSb}}
\subfloat[$K^Q_{15,18}$]{\includegraphics[width=0.66\columnwidth]{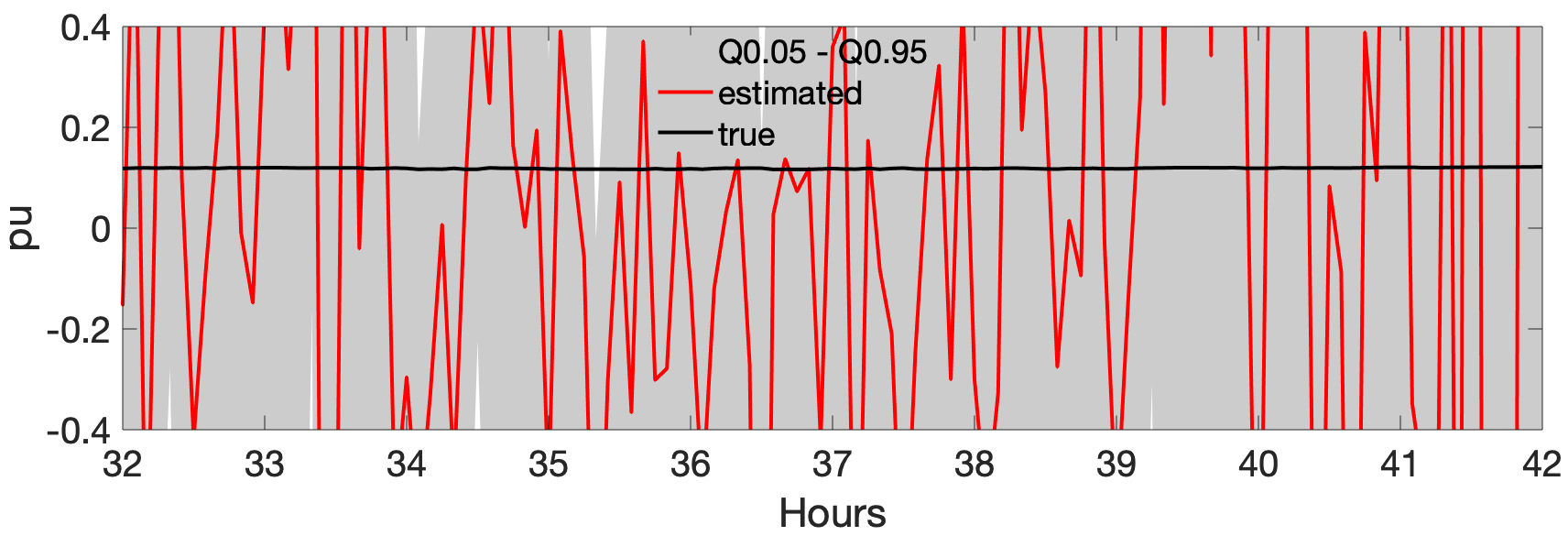}
\label{fig:LSc}}
\caption{Coefficients estimates and their uncertainty using  \textcolor{black}{the} LS.} \label{fig:LS}
\end{figure*}
\begin{figure*}[!h]
\centering
\hfil
\subfloat[$K^P_{15,15}$]{\includegraphics[width=0.66\columnwidth]{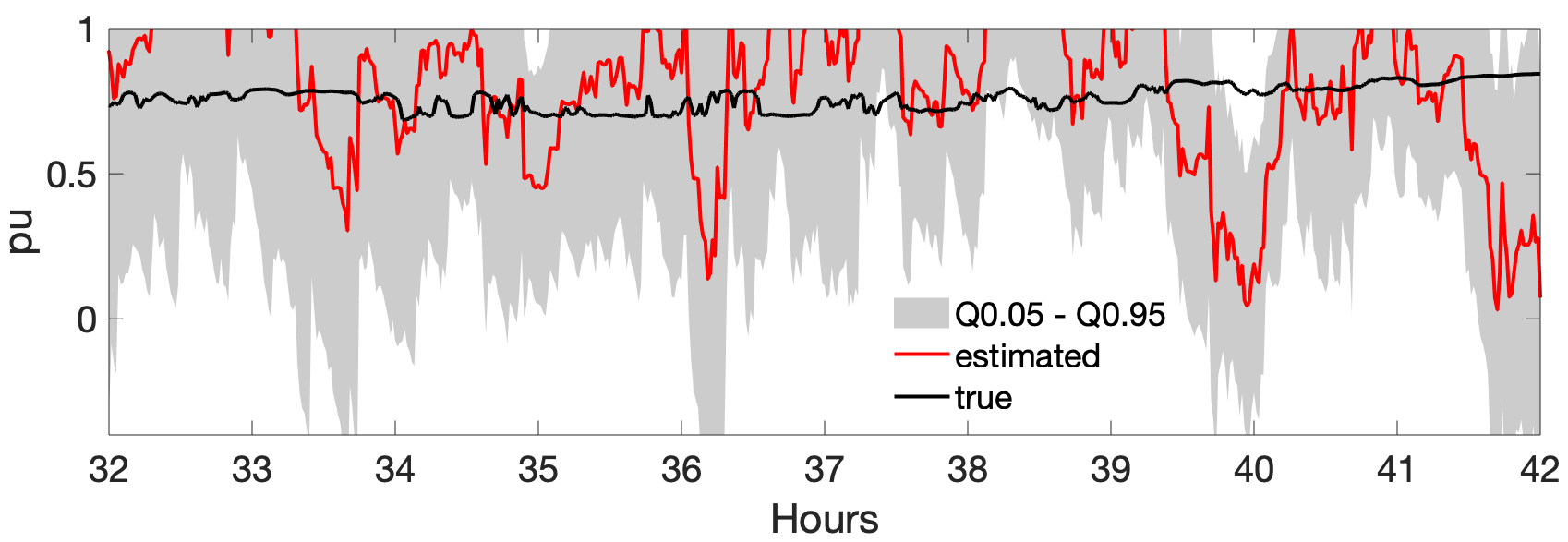}
\label{fig:RLS-Fa}}
\hfil
\subfloat[$K^P_{14,8}$]{\includegraphics[width=0.66\columnwidth]{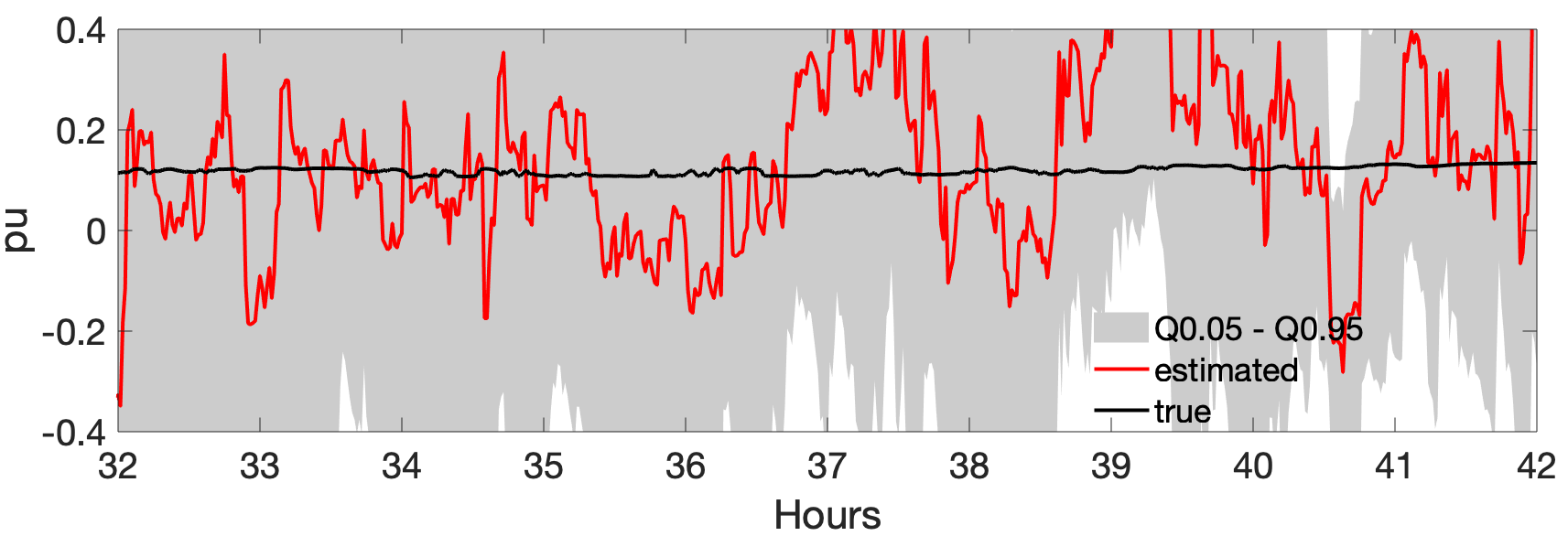}
\label{fig:RLS-Fb}}
\subfloat[$K^Q_{15,18}$]{\includegraphics[width=0.66\columnwidth]{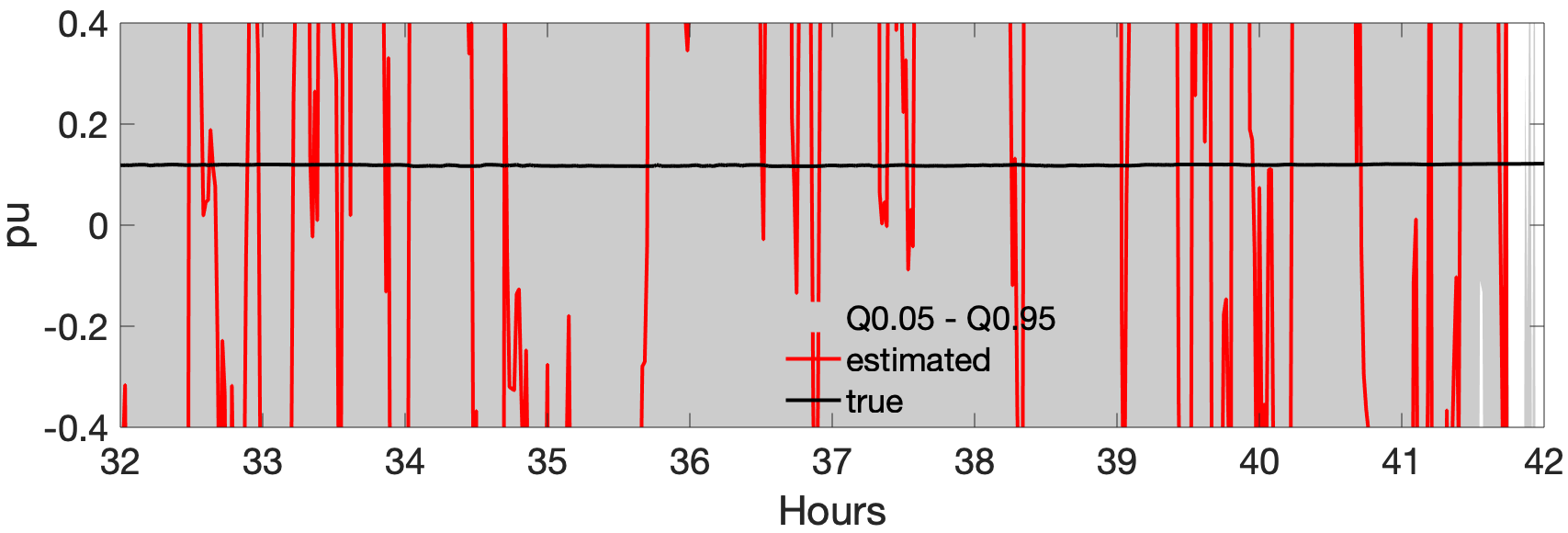}
\label{fig:RLS-Fc}}
\caption{Coefficients estimates and their uncertainty using  \textcolor{black}{the} RLS-F.} \label{fig:RLS-F}
\end{figure*}
\begin{figure*}[!h]
\centering
\subfloat[$K^P_{15,15}$]{\includegraphics[width=0.66\columnwidth]{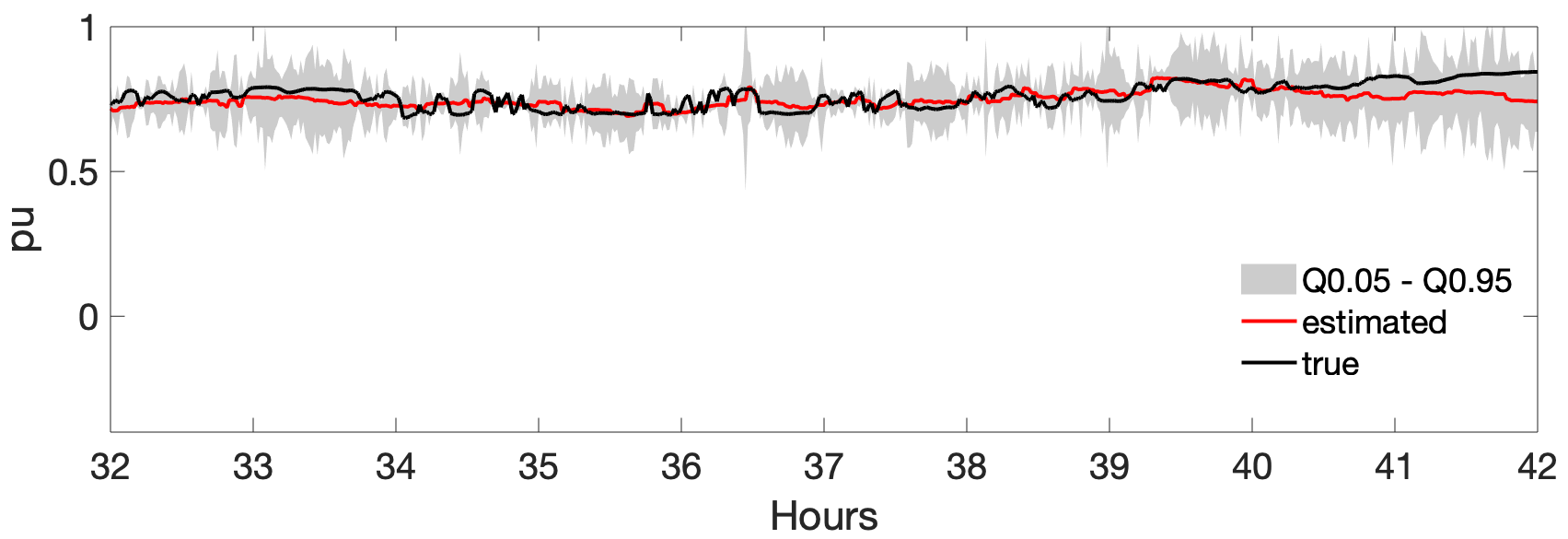}
\label{fig:RLS-CTa}}
\subfloat[$K^P_{14,8}$]{\includegraphics[width=0.66\columnwidth]{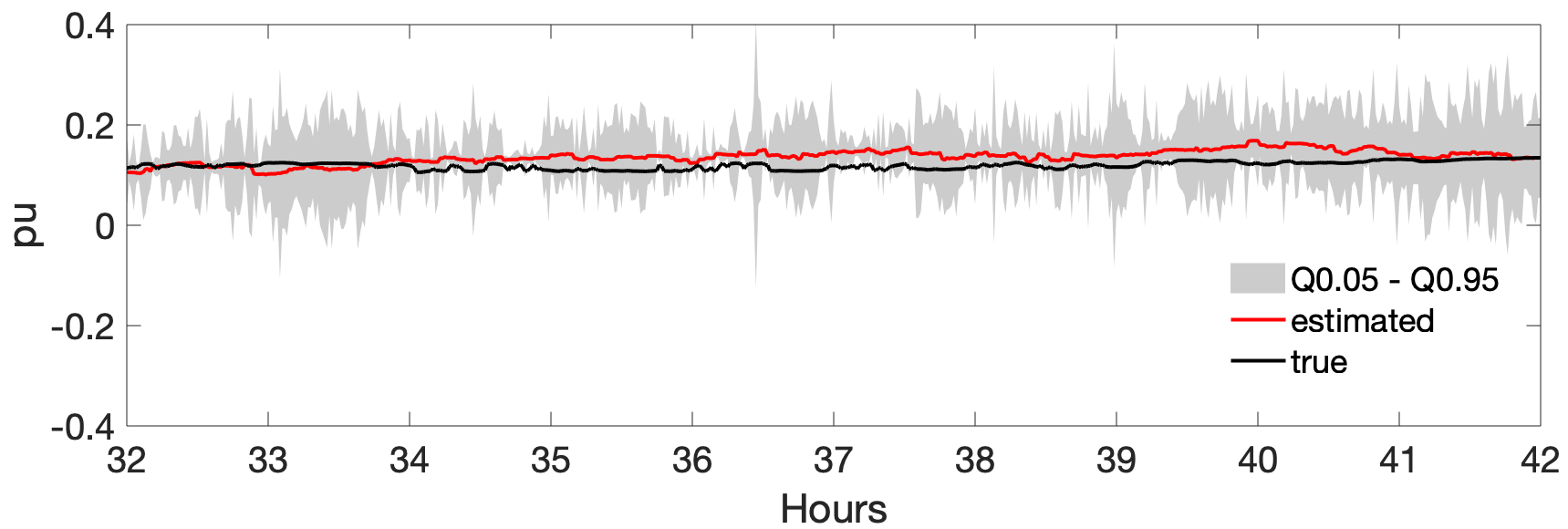}
\label{fig:RLS-CTb}}
\subfloat[$K^Q_{15,18}$]{\includegraphics[width=0.66\columnwidth]{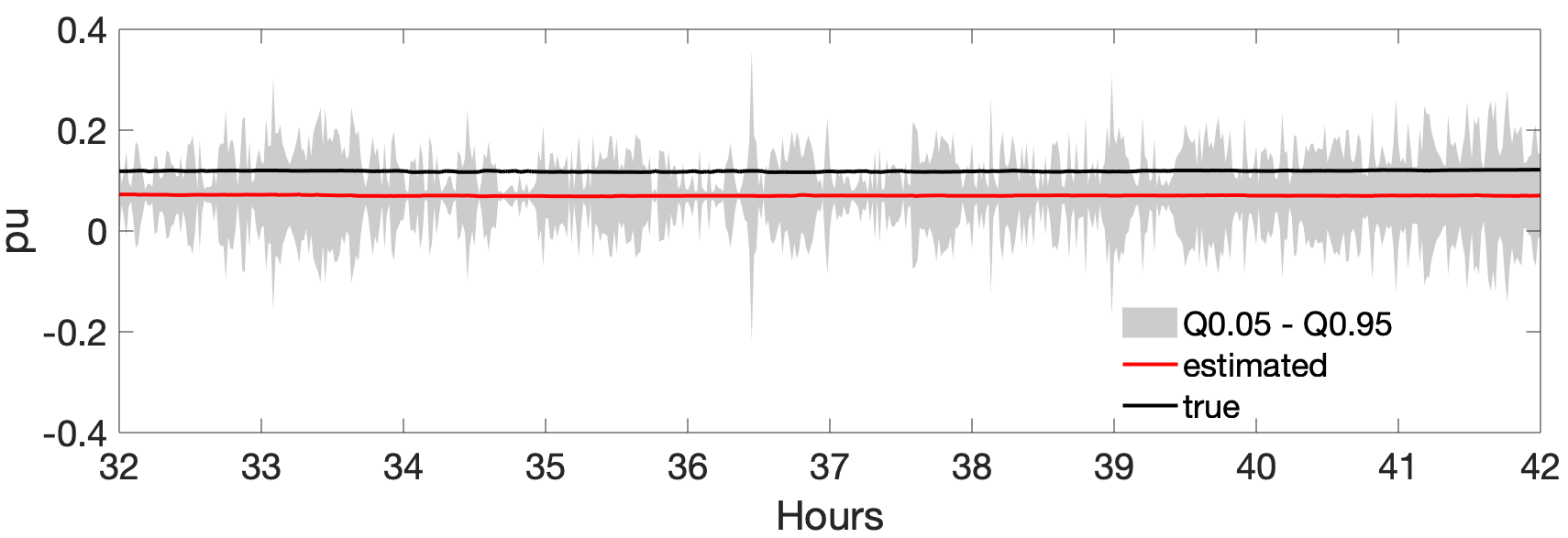}
\label{fig:RLS-CTc}}
\caption{Coefficients estimates and their uncertainty using  \textcolor{black}{the} RLS-CT.} \label{fig:RLS-CT}
\end{figure*}
\begin{figure*}[!h]
\centering
\hfil
\subfloat[$K^P_{15,15}$]{\includegraphics[width=0.66\columnwidth]{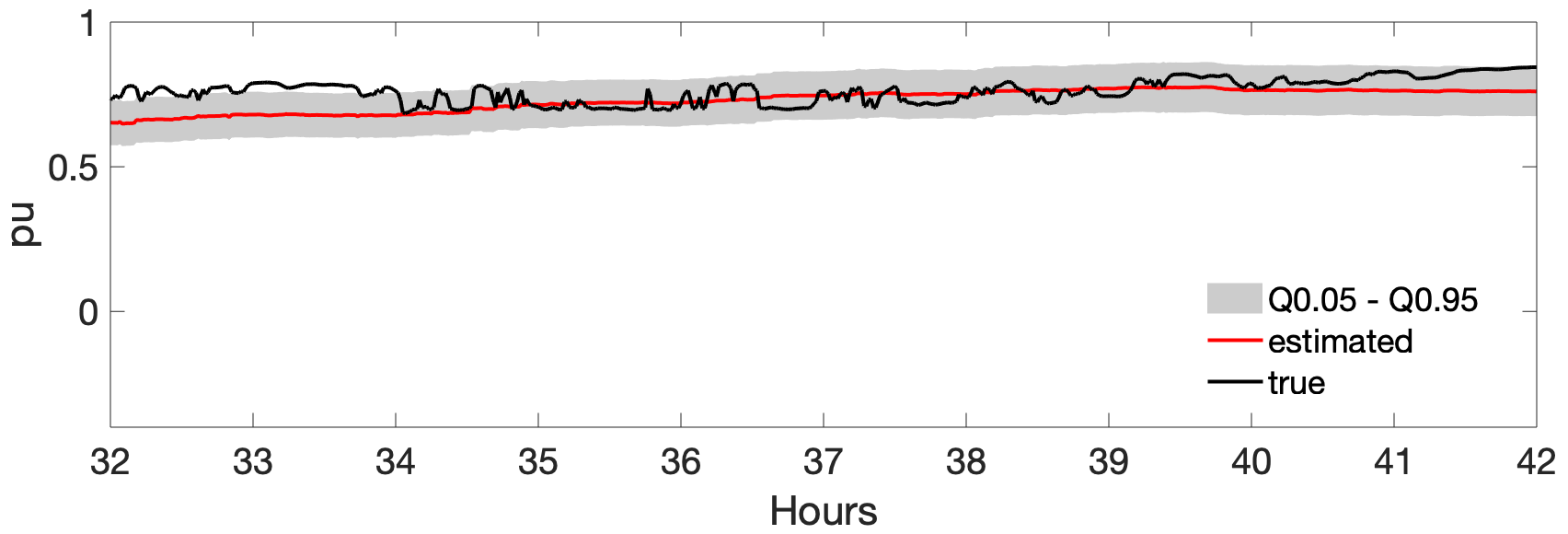}
\label{fig:RLS-SFa}}
\hfil
\subfloat[$K^P_{14,8}$]{\includegraphics[width=0.66\columnwidth]{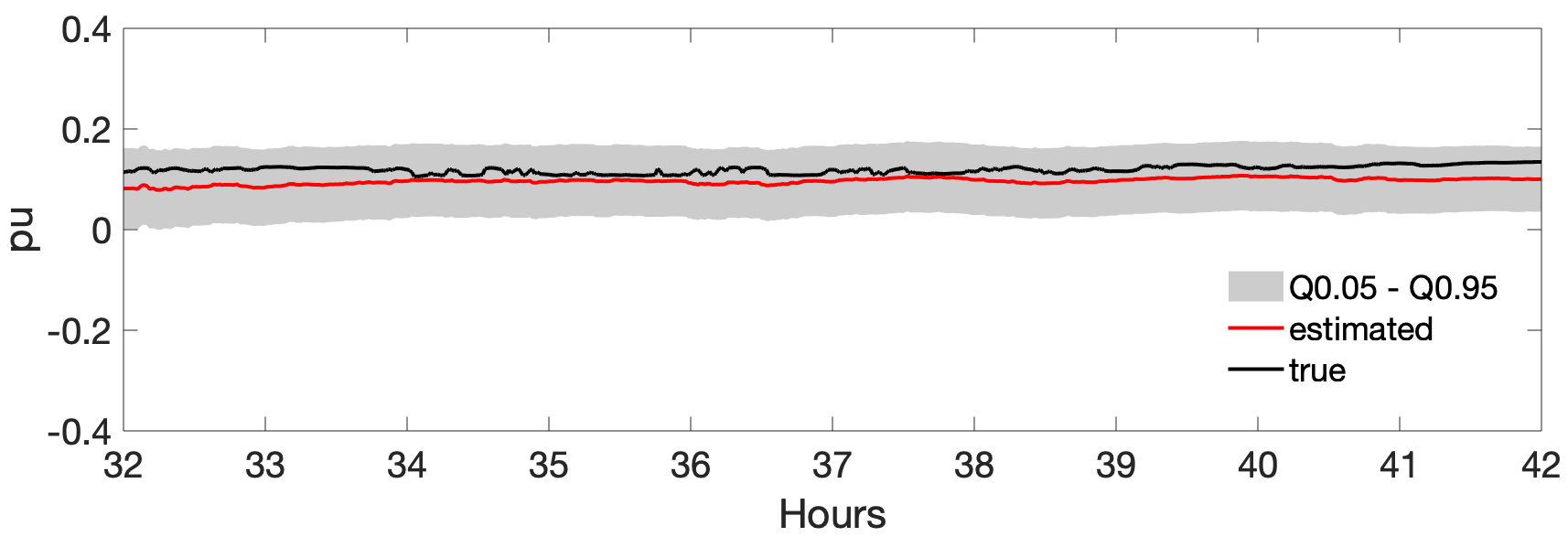}
\label{fig:RLS-SFb}}
\subfloat[$K^Q_{15,18}$]{\includegraphics[width=0.66\columnwidth]{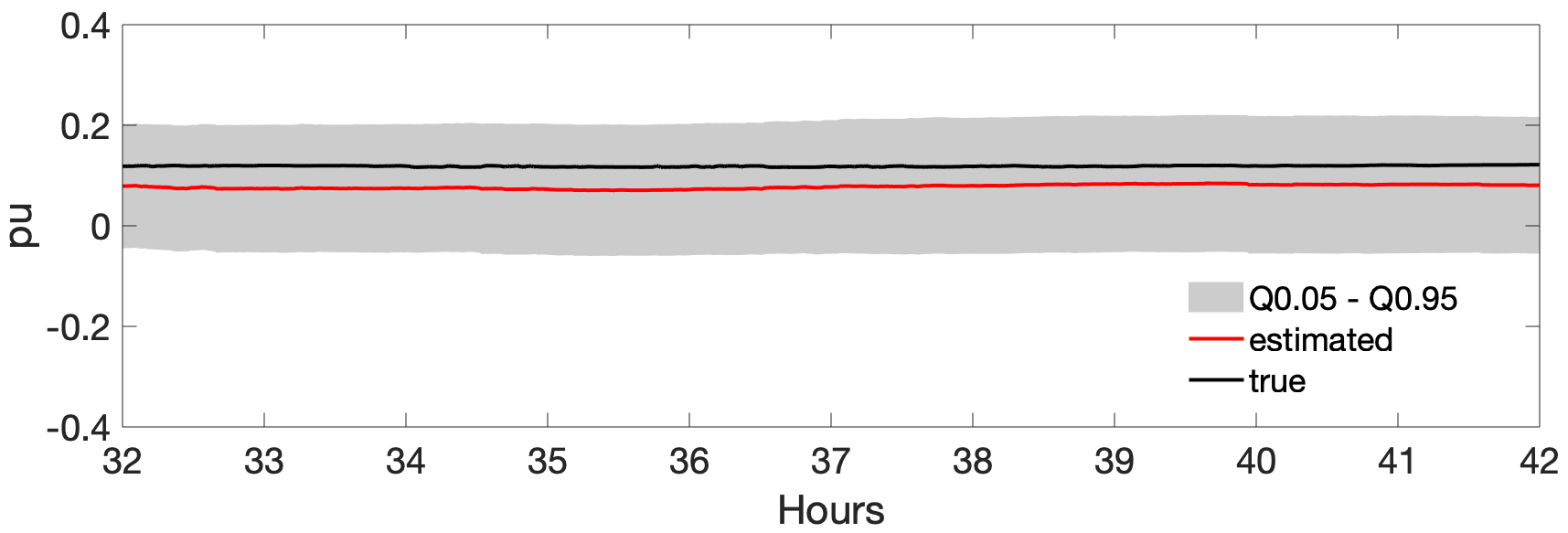}
\label{fig:RLS-SFc}}
\caption{Coefficients estimates and their uncertainty using  \textcolor{black}{the} RLS-SF.} \label{fig:RLS-SF}
\end{figure*}
\begin{figure*}[!h]
\centering
\hfil
\subfloat[$K^P_{15,15}$]{\includegraphics[width=0.66\columnwidth]{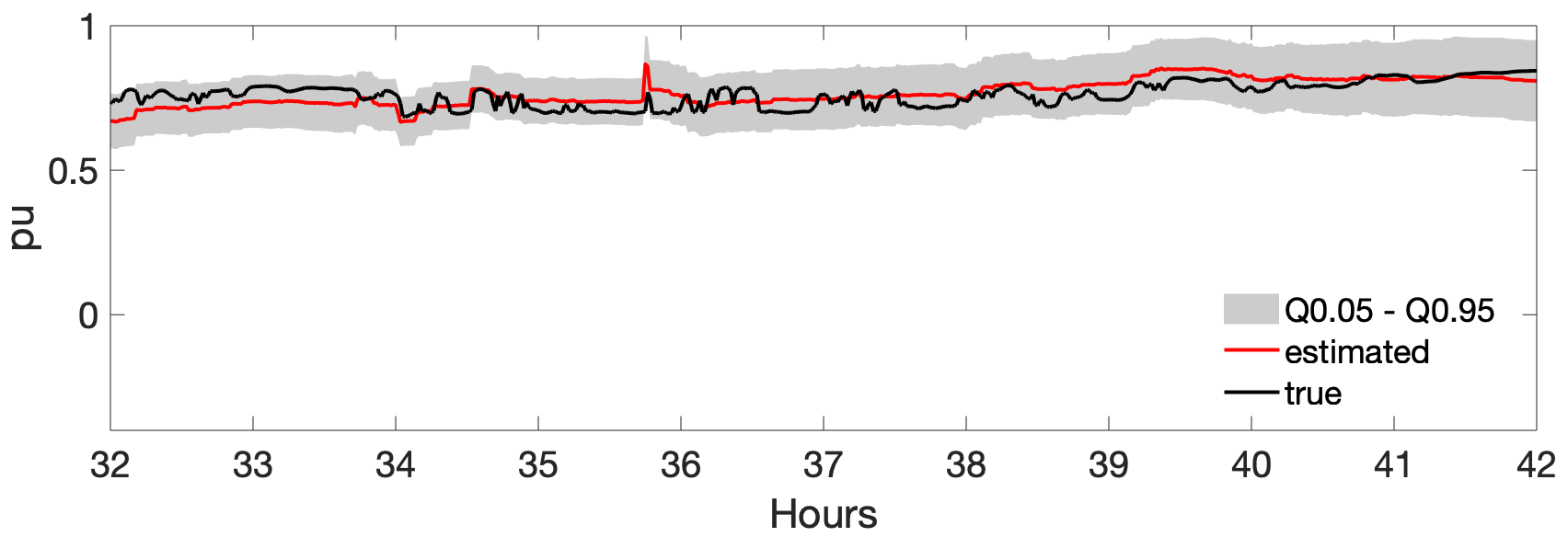}
\label{fig:RLS-DFa}}
\hfil
\subfloat[$K^P_{14,8}$]{\includegraphics[width=0.66\columnwidth]{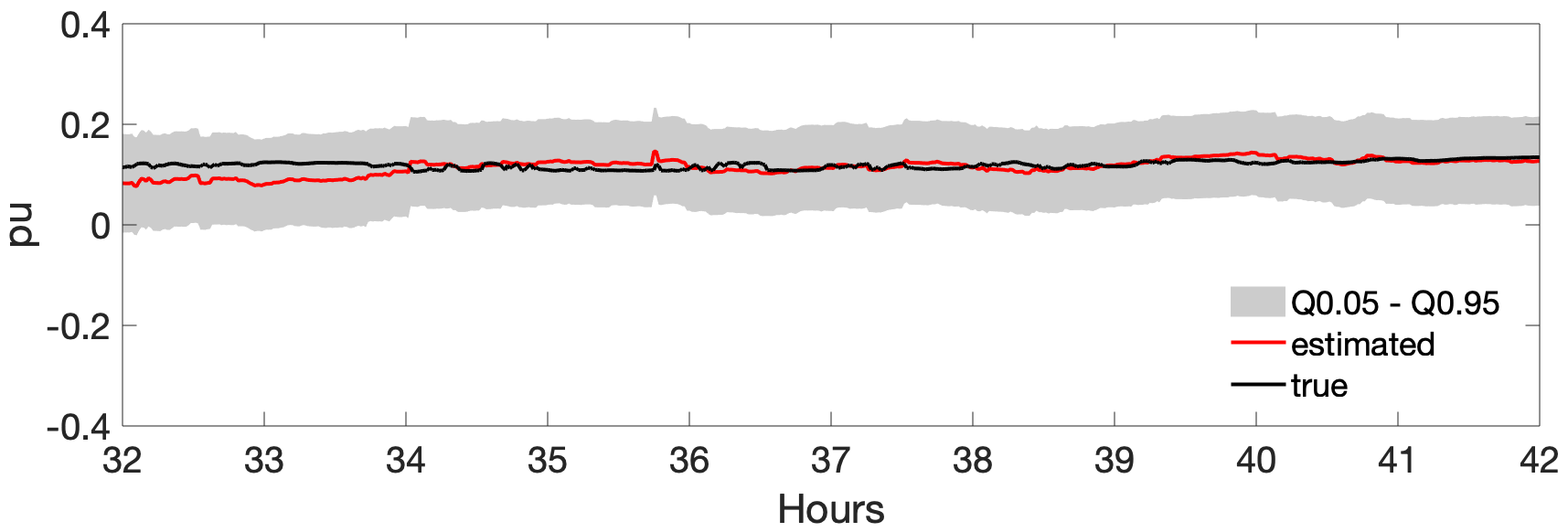}
\label{fig:RLS-DFb}}
\subfloat[$K^Q_{15,18}$]{\includegraphics[width=0.66\columnwidth]{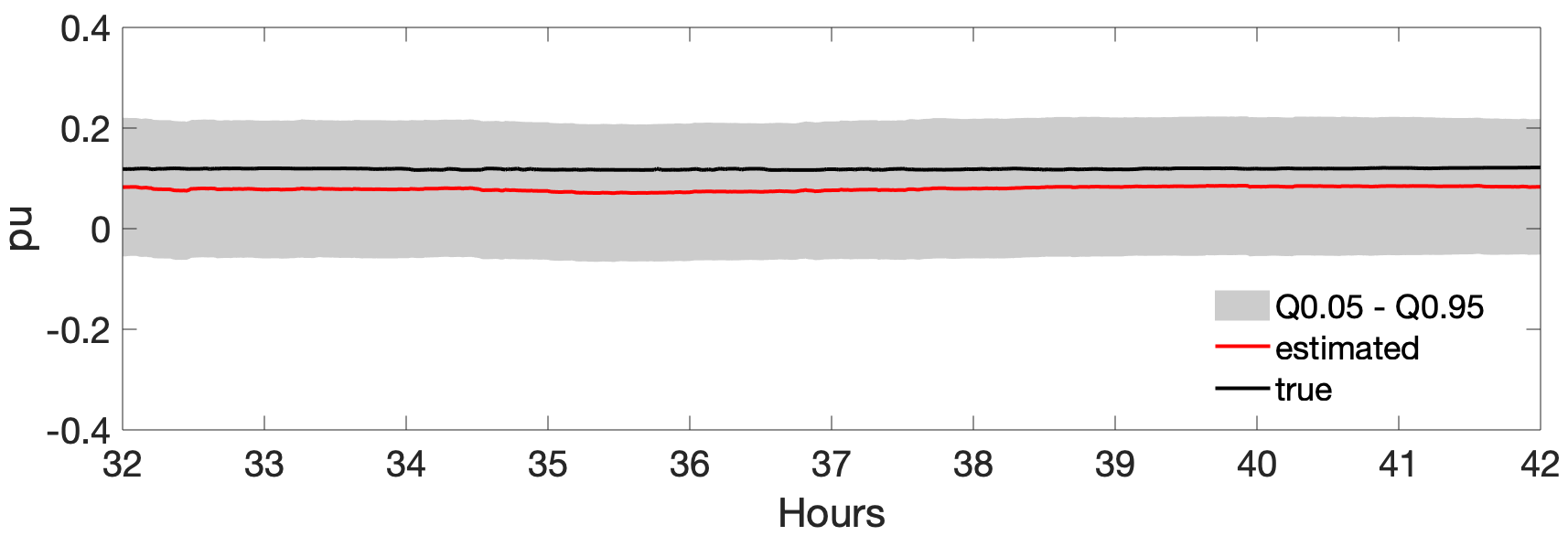}
\label{fig:RLS-DFc}}
\caption{Coefficients estimates and their uncertainty using  \textcolor{black}{the} RLS-DF.} \label{fig:RLS-DF}
\end{figure*}
\begin{table*}[!h]
    \centering
    \caption{Performance comparison of different estimation techniques for $K^P_{15,15}$ with different IT classes.}
    \begin{tabular}{|c|c|c||c|c||c|c|}
    \hline
         & \multicolumn{2}{|c||}{\bf{}IT 0.2} & \multicolumn{2}{|c||}{\bf{}IT 0.5} & \multicolumn{2}{|c|}{\bf{}IT 1.0}\\
         \hline
         \bf{Method} & \bf{RMSE} & \bf{PICP-PINAW-CWC}& \bf{RMSE} & \bf{PICP-PINAW-CWC} & \bf{RMSE} & \bf{PICP-PINAW-CWC}  \\
         \hline
         \bf{LS} & 0.89 & 0.14-0.36-0.41 & 0.94 & 0.03-0.35-0.37 & 0.98 & 0.04-0.41-0.43\\
         \hline
         \bf{RLS-F} & 0.46 & 0.99-1.05-1.85 & 0.39 & 0.93-1.45-2.70 & 0.43 & 0.86-1.32-2.38\\
         \hline
         \bf{RLS-CT} & 0.16 & 0.18-0.11-0.11 & 0.05 & 0.87-0.19-0.34 & 0.05 & 0.88-0.23-0.42\\
         \hline
         \bf{RLS-SF} & 0.05 & 1.00-0.42-0.42 & 0.07 & 0.83-0.19-0.35 & 0.06 & 0.89-0.18-0.33\\
         \hline
         \bf{RLS-DF} & 0.05 & 1.00-0.47-0.47 & 0.05 & 0.99-0.24-0.24 & 0.06 & 0.99-0.26-0.26\\
         \hline
    \end{tabular}
    \label{tab:estimation_comparison}
\end{table*}
\subsection{Estimation results}
We estimate $\mathbf{K}^P, \mathbf{K}^Q$ for the nodes where the controllable units (i.e., PV generation units) are connected. 
The estimation results using the measurements corresponding to IT 1.0 are presented below.
The estimated coefficients are shown for the 2nd day with peak PV production during 32 - 42 hours (potentially causing over-voltages). For performance comparison among different schemes, we report the estimations for LS, RLS-F, RLS-CT, RLS-SF, and RLS-DF as defined in Sec.~\ref{sec:estimation_tech}. ``LS" solves the LS algorithm and uses the measurements from last 5-minutes (sampled at 1-second) to estimate the sensitivity coefficients. For the methods based on the RLS, the first-day measurements (0 - 24 hours) are used to compute initial estimates (offline-LS). Then, they are updated each 5-minutes with the last measurements in a recursive way. The forgetting factor $\mu = 0.85$ is used in the simulations.

Figures~\ref{fig:LS}-\ref{fig:RLS-DF} shows the estimations and prediction intervals with confidence interval coverage of 99\%. For the sake of brevity, we show only three coefficients which are $K^P_{15,15}, K^P_{14,8},K^Q_{15,18}$ in Fig~\ref{fig:LS}-\ref{fig:RLS-DF}~(a), (b) and (c) respectively. The plots in red show the estimated coefficients, and the gray area their \textcolor{black}{corresponding} uncertainty. The black line shows the true coefficients. As observed from the plot, LS fails in \textcolor{black}{reliably} estimating the coefficients and suffers from biases and large variances. \textcolor{black}{The} RLS-F exhibits large uncertainty on the estimates. This is due to the windup problem in the covariance matrix, as reported in \cite{parkum1992recursive}. RLS-CT, RLS-SF, and RLS-DF \textcolor{black}{do fix the} windup problem using the strategies described in Sec.~\ref{sec:estimation_tech}. However, the RLS-CT fails \textcolor{black}{to reliably estimate} for the coefficient $K^Q_{15,18}$. RLS-SF and RLS-DF show similar performances. However, the former fails to cover the true coefficient during 32 - 34 hours for $K^P_{15,15}$.

To have a proper comparison, we report \textcolor{black}{the} RMSE and \textcolor{black}{the} PICP-PINAW-CWC in Table~\ref{tab:estimation_comparison} for the coefficient $K^P_{15,15}$ for the same duration (32 - 42 hours) using different methods and with measurements characterized by other IT classes. From \textcolor{black}{such a} comparison, it can be observed \textcolor{black}{that the} RLS-DF performs the best \textcolor{black}{with respect to} all the metrics, i.e., it has the lowest RMSE and highest coverage.
From Table~\ref{tab:estimation_comparison}, it can be observed that for all the estimation techniques, the RMSE increases for increasing measurement noise. \textcolor{black}{The} RLS-DF has full PICP coverage for all the IT classes, whereas  \textcolor{black}{the} RLS-SF has slightly  \textcolor{black}{lower} PICP for IT 0.5 and IT 1.0. From the comparison, it can be concluded that RLS-DF is the dominant estimation method.
\subsection{Control results}
In the following, we present the voltage control results. We control all three PV plants using the robust and non-robust approaches \textcolor{black}{described in Sec.~\ref{sec:v_control}}. The objective is to restrict the voltage magnitudes within \textcolor{black}{the} bounds 0.97 - 1.03 pu. 
\textcolor{black}{We show the results only using the dominant estimation schemes i.e., RLS-SF and RLS-DF from the last analysis. In this setup, the previous day measurements (0-24 hrs) are used to compute initial estimates (offline-LS) for the RLS. Then,} \textcolor{black}{the} online-RLS refines these estimates every 5 minutes using the last measurement. 
\textcolor{black}{The latest estimated coefficients and their uncertainties are then used for the} voltage control.

\begin{figure}[!h]
\centering
\subfloat[]{\includegraphics[width=1\columnwidth]{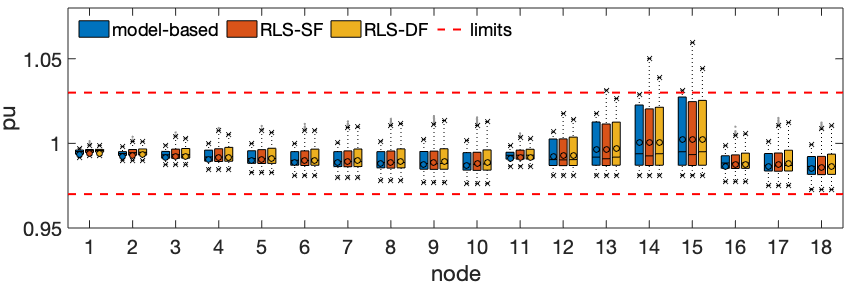}
\label{fig:VnonRObust}}
\hfil
\subfloat[]{\includegraphics[width=1\columnwidth]{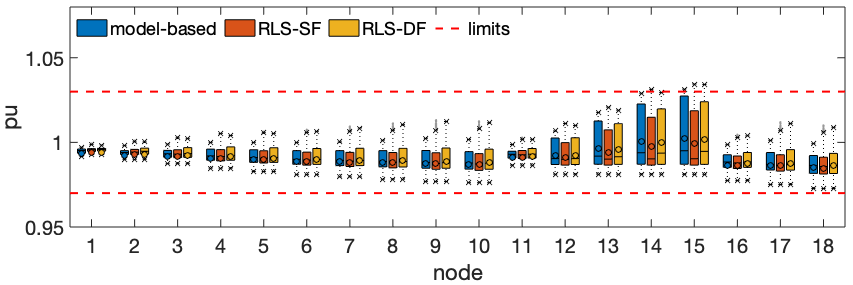}
\label{fig:Vrobust}}
\caption{Distribution of \textcolor{black}{daily} nodal voltage magnitudes using (a) non-robust and (b) robust voltage control.} \label{fig:RobustvsNonRobust}
\end{figure}

\paragraph{Non-robust voltage control} 
Fig.~\ref{fig:VnonRObust} \textcolor{black}{compares the daily boxplot post-control nodal voltage magnitudes for all the nodes using \textcolor{black}{the} non-robust control.} The performance is also compared against \emph{model-based} method, i.e., when the true sensitivity coefficients are known.
As clear from the comparison, the non-robust control fails to restrict the voltage magnitudes of nodes 14 and 15 within imposed bounds by a large margin, irrespective of the estimation techniques. It should be noted that even the dominant estimation method (i.e., RLS-DF) fails to respect the upper voltage constraint in non-robust control. 
\paragraph{Robust voltage control}
Figure~\ref{fig:Vrobust} \textcolor{black}{compares the daily boxplot post-control nodal voltage magnitudes for all the nodes using the robust voltage control.}
As it can be seen, robust voltage control succeeds in reducing voltage violations. Robust voltage control using estimates from RLS-SF and RLS-DF perform similarly to the model-based controls (i.e., the maximum voltage magnitude is near the upper bound).

\begin{figure}[!h]
\centering
\subfloat[]{\includegraphics[width=1\columnwidth]{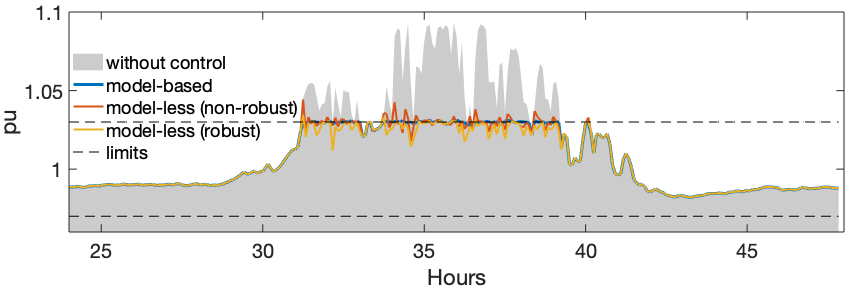}
\label{fig:VRLSDF}}
\hfil
\subfloat[]{\includegraphics[width=1\columnwidth]{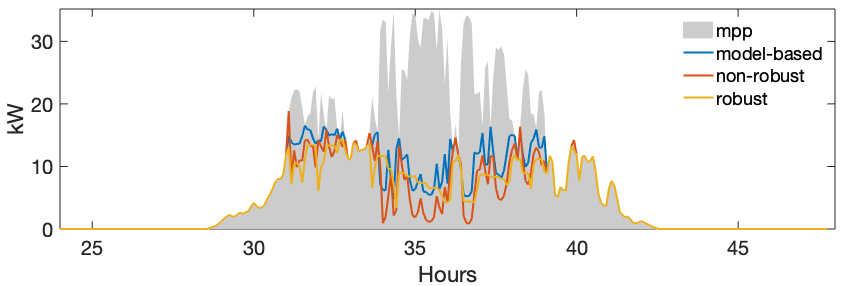}
\label{fig:PVRLSDF}}
\hfil
\subfloat[]{\includegraphics[width=1\columnwidth]{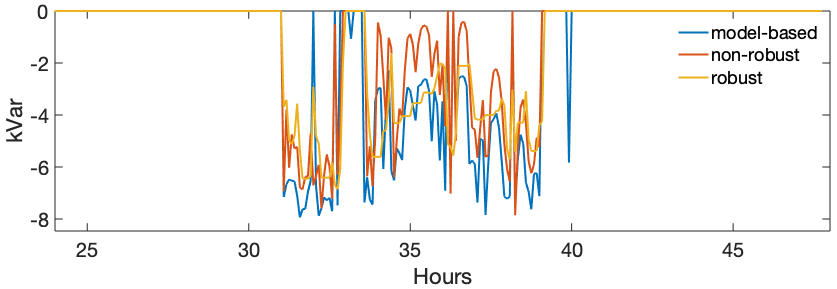}
\label{fig:PVqRLSDF}}
\caption{Control results using RLS-DF for robust, non-robust and model-based controls: (a) voltage magnitude, (b) active power and (c) reactive power for node 15.} \label{fig:VControl_RLS-DF}
\end{figure}

Figure~\ref{fig:VControl_RLS-DF} shows the control results for \textcolor{black}{the} RLS-DF, comparing model-less robust and non-robust methods against model-based control. Fig.~\ref{fig:VRLSDF} shows the voltage of node 15 under different control schemes. It can be observed that model-less robust control keeps the voltage within the imposed upper bound and close to the model-based approach, whereas the non-robust method has higher voltage violations.  Fig.~\ref{fig:PVRLSDF} shows the curtailed PV generation for node 15, and it can be seen that model-less controls curtail more than model-based control. The model-based control curtails 86.5~kWh out of 210~kWh (MPP), whereas model-less non-robust and robust schemes curtailed 106 and 104~kWh respectively. This is because they provide a conservative solution to avoid voltage violations. Although non-robust scheme curtails more, it failed to respect the voltage bounds due to inaccurate reactive power actuation.
Finally, Fig.~\ref{fig:PVqRLSDF} shows the reactive power injections in three cases. Model-based and robust control follow a similar pattern, whereas non-robust provides less reactive power during the middle of the day.
\paragraph{Performance with measurement noise}
We also present a performance comparison when robust or non-robust control is coupled with different estimation techniques for different IT classes of measurement noise. The results are summarised in Table~\ref{tab:VmaxComparison} \textcolor{black}{resulting in different and the following observations}: (i) non-robust control always results in voltage violations, even when the measurement noise is minimum; in contrast, robust control achieves negligible violations; (ii) RLS-SF and RLS-DF-based robust control performed the best \textcolor{black}{with respect to} maximum voltage violations, irrespective of \textcolor{black}{the} IT class. 


\begin{table*}[!h]
    \centering
    \caption{Performance comparison of different voltage control methods: maximum nodal voltage magnitude.}
    \begin{tabular}{|c|c|c||c|c||c|c|}
    \hline
         & \multicolumn{2}{|c||}{\bf{}IT 0.2} & \multicolumn{2}{|c||}{\bf{}IT 0.5} & \multicolumn{2}{|c|}{\bf{}IT 1.0}\\
         \hline
         \bf{Method} & \bf{non-Robust} & \bf{Robust}& \bf{non-Robust} & \bf{Robust} & \bf{non-Robust} & \bf{Robust}  \\
         \hline
         \bf{Model-based}  & \multicolumn{6}{|c|}{1.031}\\
         \hline
         \bf{RLS-SF (model-less)} & 1.041 & 1.032 & 1.043 & 1.032 & 1.059 & 1.034\\
         \hline
         \bf{RLS-DF (model-less)} & 1.036 & 1.031 & 1.043 & 1.034 & 1.045 & 1.034\\
         \hline
    \end{tabular}
    \label{tab:VmaxComparison}
\end{table*}

\section{Conclusion}\label{sec:conclusion}
This work proposed a model-less robust voltage control scheme accounting for the uncertainty on the sensitivity coefficients which are estimated from measurements. The control framework consisted of two stages: in the first, voltage sensitivity coefficients and their uncertainties are estimated using the measurements of nodal voltage magnitudes and active and reactive power. In the second stage, these estimated coefficients and their uncertainties are used \textcolor{black}{by} for the robust voltage control problems.

The voltage sensitivity coefficients are estimated using a recursive estimation algorithm, where \textcolor{black}{the} LS is solved offline to provide a rough estimate of the coefficient using a large number of historical measurements. Then, RLS is used to \textcolor{black}{refine such a preliminary estimation by} using the most recent measurements. The work also compares different forgetting schemes in online RLS estimation. We incorporate the uncertainty of the estimated coefficient to formulate a robust voltage control. 

The scheme is validated for controlling active/reactive power injections from distributed PV generation units connected to \textcolor{black}{the} CIGRE low-voltage benchmark network. The results show that the non-robust voltage controls fail to \textcolor{black}{satisfy} the voltage constraint (i.e., when uncertainty on the estimated coefficients are not accounted for). The proposed robust control scheme respects the voltage control limit even in the highest instrument class. The performance comparison with respect to different estimation schemes shows that an online estimation scheme with directional forgetting performs the best.

\bibliographystyle{IEEEtran}
\bibliography{biblio.bib}

\begin{thebibliography}{10}
\providecommand{\url}[1]{#1}
\csname url@samestyle\endcsname
\providecommand{\newblock}{\relax}
\providecommand{\bibinfo}[2]{#2}
\providecommand{\BIBentrySTDinterwordspacing}{\spaceskip=0pt\relax}
\providecommand{\BIBentryALTinterwordstretchfactor}{4}
\providecommand{\BIBentryALTinterwordspacing}{\spaceskip=\fontdimen2\font plus
\BIBentryALTinterwordstretchfactor\fontdimen3\font minus
  \fontdimen4\font\relax}
\providecommand{\BIBforeignlanguage}[2]{{%
\expandafter\ifx\csname l@#1\endcsname\relax
\typeout{** WARNING: IEEEtran.bst: No hyphenation pattern has been}%
\typeout{** loaded for the language `#1'. Using the pattern for}%
\typeout{** the default language instead.}%
\else
\language=\csname l@#1\endcsname
\fi
#2}}
\providecommand{\BIBdecl}{\relax}
\BIBdecl

\bibitem{guide2004voltage}
{{Power Quality Application Guide}}, ``Voltage disturbances,'' \emph{Standard
  EN}, vol. 50160, 2004.

\bibitem{CIGREREF}
{{CIGRE' Task Force C6.04.02}}, ``Benchmark systems for network integration of
  renewable and distributed energy resources,'' Cigre' International Council on
  large electric systems, Tech. Rep., July 2009.

\bibitem{IEEE_practice}
{{IEEE Std} 1159-2009}, ``Ieee recommended practice for monitoring electric
  power quality,'' (Revision of IEEE Std 1159-1995), Tech. Rep., June 2009.

\bibitem{hatziargyrioucigre}
N.~Hatziargyriou, J.~Amantegui, B.~Andersen, M.~Armstrong, P.~Boss, B.~Dalle,
  G.~de~Montravel, A.~Negri, C.~A. Nucci, and P.~Southwell, ``{CIGRE WG
  “Network of the Future”},'' June 2011.

\bibitem{cigre2011c6}
W.~CIGR{\'E}, ``C6. 11,“development and operation of active distribution
  networks,” cigr{\'e}, paris, idf,'' FR, Tech. Rep. 457, Tech. Rep., 2011.

\bibitem{pilo2012planning}
F.~Pilo, S.~Jupe, F.~Silvestro, K.~El~Bakari, C.~Abbey, G.~Celli, J.~Taylor,
  A.~Baitch, and C.~Carter-Brown, ``Planning and optimisation of active
  distribution systems-an overview of cigre working group c6. 19 activities,''
  2012.

\bibitem{christakou2015real}
K.~Christakou, ``Real-time optimal controls for active distribution networks:
  from concepts to applications,'' Ph.D. dissertation, Ph. D. dissertation,
  Dept. Information and communications, Univ. {\'E}cole~…, 2015.

\bibitem{agalgaonkar2013distribution}
Y.~P. Agalgaonkar, B.~C. Pal, and R.~A. Jabr, ``Distribution voltage control
  considering the impact of pv generation on tap changers and autonomous
  regulators,'' \emph{IEEE Transactions on Power Systems}, vol.~29, no.~1, pp.
  182--192, 2013.

\bibitem{gupta2020grid}
R.~K. Gupta, F.~Sossan, and M.~Paolone, ``Grid-aware distributed model
  predictive control of heterogeneous resources in a distribution network:
  Theory and experimental validation,'' \emph{IEEE Trans. Energy Conv.}, 2020.

\bibitem{su2019augmented}
H.~Su, P.~Li, X.~Fu, L.~Yu, and C.~Wang, ``Augmented sensitivity estimation
  based voltage control strategy of active distribution networks with pmu
  measurement,'' \emph{IEEE Access}, vol.~7, pp. 44\,987--44\,997, 2019.

\bibitem{carpita2019low}
M.~Carpita, A.~Dassatti, M.~Bozorg, J.~Jaton, S.~Reynaud, and O.~Mousavi, ``Low
  voltage grid monitoring and control enhancement: The grideye solution,'' in
  \emph{2019 ICCEP}.\hskip 1em plus 0.5em minus 0.4em\relax IEEE, 2019, pp.
  94--99.

\bibitem{da2019data}
E.~L. da~Silva, A.~M.~N. Lima, M.~B. de~Rossiter~Corr{\^e}a, M.~A. Vitorino,
  and L.~T. Barbosa, ``Data-driven sensitivity coefficients estimation for
  cooperative control of pv inverters,'' \emph{IEEE Transactions on Power
  Delivery}, vol.~35, no.~1, pp. 278--287, 2019.

\bibitem{nowak2020measurement}
S.~Nowak, Y.~C. Chen, and L.~Wang, ``Measurement-based optimal der dispatch
  with a recursively estimated sensitivity model,'' \emph{IEEE Transactions on
  Power Systems}, vol.~35, no.~6, pp. 4792--4802, 2020.

\bibitem{zhang2017locally}
J.~Zhang, Z.~Wang, X.~Zheng, L.~Guan, and C.~Chung, ``Locally weighted ridge
  regression for power system online sensitivity identification considering
  data collinearity,'' \emph{IEEE Trans. Power Syst.}, vol.~33, no.~2, pp.
  1624--1634, 2017.

\bibitem{zhang2017noise}
J.~Zhang, C.~Chung, and L.~Guan, ``Noise effect and noise-assisted ensemble
  regression in power system online sensitivity identification,'' \emph{IEEE
  Trans. Ind. Info.}, vol.~13, no.~5, pp. 2302--2310, 2017.

\bibitem{bertsimas2011theory}
D.~Bertsimas, D.~B. Brown, and C.~Caramanis, ``Theory and applications of
  robust optimization,'' \emph{SIAM review}, vol.~53, no.~3, pp. 464--501,
  2011.

\bibitem{soderstrom1989system}
T.~S{\"o}derstr{\"o}m and P.~Stoica, \emph{System identification}.\hskip 1em
  plus 0.5em minus 0.4em\relax Prentice-Hall International, 1989.

\bibitem{mills1991internet}
D.~L. Mills, ``Internet time synchronization: the network time protocol,''
  \emph{IEEE Trans. comm.}, vol.~39, no.~10, pp. 1482--1493, 1991.

\bibitem{parkum1992recursive}
J.~Parkum, N.~K. Poulsen, and J.~Holst, ``Recursive forgetting algorithms,''
  \emph{International Journal of Control}, vol.~55, no.~1, pp. 109--128, 1992.

\bibitem{vahidi2005recursive}
A.~Vahidi, A.~Stefanopoulou, and H.~Peng, ``Recursive least squares with
  forgetting for online estimation of vehicle mass and road grade: theory and
  experiments,'' \emph{Vehicle System Dynamics}, vol.~43, no.~1, pp. 31--55,
  2005.

\bibitem{fortescue1981implementation}
T.~Fortescue, L.~S. Kershenbaum, and B.~E. Ydstie, ``Implementation of
  self-tuning regulators with variable forgetting factors,'' \emph{Automatica},
  vol.~17, no.~6, pp. 831--835, 1981.

\bibitem{cao1999novel}
L.~Cao and H.~M. Schwartz, ``A novel recursive algorithm for directional
  forgetting,'' in \emph{Proceedings of the 1999 American Control Conference
  (Cat. No. 99CH36251)}, vol.~2.\hskip 1em plus 0.5em minus 0.4em\relax IEEE,
  1999, pp. 1334--1338.

\bibitem{bittanti1990exponential}
S.~Bittanti, P.~Bolzern, and M.~Campi, ``Exponential convergence of a modified
  directional forgetting identification algorithm,'' \emph{Systems \& Control
  Letters}, vol.~14, no.~2, pp. 131--137, 1990.

\bibitem{christakou2017voltage}
K.~Christakou, M.~Paolone, and A.~Abur, ``Voltage control in active
  distribution networks under uncertainty in the system model: A robust
  optimization approach,'' \emph{IEEE Trans. Smart Grid}, vol.~9, no.~6, pp.
  5631--5642, 2017.

\bibitem{bertsimas2004price}
D.~Bertsimas and M.~Sim, ``The price of robustness,'' \emph{Operations
  research}, vol.~52, no.~1, pp. 35--53, 2004.

\bibitem{IT_V}
{\textit{Instrument Transformers:}}, ``Additional requirements for electronic
  voltage transformers,'' \emph{{Standard IEC}}, pp. 61\,869--8, 2011.

\bibitem{IT_C}
{\it{Instrument Transformers:}}, ``Additional requirements for electronic
  current transformers,'' \emph{{Standard IEC}}, pp. 61\,869--7, 2014.

\bibitem{khosravi2013prediction}
A.~Khosravi, S.~Nahavandi, and D.~Creighton, ``Prediction intervals for
  short-term wind farm power generation forecasts,'' \emph{IEEE Transactions on
  sustainable energy}, vol.~4, no.~3, pp. 602--610, 2013.

\end{thebibliography}
\end{document}